\documentclass[submission,copyright,creativecommons]{eptcs}
\providecommand{\event}{FMAS 2026}

\usepackage{iftex}
\ifpdftex
  \usepackage[T1]{fontenc}
\fi

\newcommand{\doi}[1]{doi:\href{https://doi.org/#1}{\nolinkurl{#1}}}
\newcommand{\eprint}[1]{arXiv:\href{https://arxiv.org/abs/#1}{\nolinkurl{#1}}}

\usepackage{mdframed}
\usepackage{microtype}
\usepackage{color}
\usepackage{amsmath}
\usepackage{amssymb}
\usepackage{amsfonts}
\usepackage{algorithmic}
\usepackage{textcomp}
\usepackage[export]{adjustbox}
\usepackage{booktabs} 
\usepackage{commath}
\usepackage{mathtools}
\usepackage{listings}
\usepackage{ifthen}
\usepackage{xurl}
\usepackage{float}
\usepackage{xstring}
\usepackage{tikz}
\usepackage{tikz-qtree}
\usepackage{xfp}
\usepackage{syntax}
\usepackage{caption}
\usepackage{newfloat} 
\usepackage{tabularx}

\usepackage{todonotes}

\usepackage{xspace}

\usepackage{xcolor}
\newif\ifshowrevisions
\showrevisionstrue          

\definecolor{RevisionColor}{HTML}{0B5FA5}   
\definecolor{RestoredColor}{HTML}{0F7A6B}   
\definecolor{RevNoteColor}{HTML}{B25000}    

\ifshowrevisions
  
  \long\def\restored#1{{\color{RestoredColor}#1}}
  \long\def\revnote#1{{\color{RevNoteColor}\small[\textbf{Note:} #1]}}
\else
  
  \long\def\restored#1{#1}
  \long\def\revnote#1{}
\fi

\newboolean{IncludeAppendix}
\setboolean{IncludeAppendix}{true}

\newcommand{\BehaviorTree}{\mathit{BT}}
\newcommand{\Invalid}{\mathit{I}}
\newcommand{\Failure}{\mathit{F}}
\newcommand{\Running}{\mathit{R}}
\newcommand{\Success}{\mathit{S}}

\newcommand{\FailureTikz}{\texttt{\itshape{F}}}
\newcommand{\RunningTikz}{\texttt{\itshape{R}}}
\newcommand{\SuccessTikz}{\texttt{\itshape{S}}}

\definecolor{BehaviorTreeBlackboardColor}{HTML}{0000FF} 
\definecolor{BehaviorTreeEnvironmentColor}{HTML}{FF8C00} 
\definecolor{NeuralVarColor}{HTML}{A0522D}

\newcommand{\EnvVar}[1]{\textcolor{BehaviorTreeEnvironmentColor}{#1}}
\newcommand{\BlVar}[1]{\textcolor{BehaviorTreeBlackboardColor}{#1}}

\newcommand{\EnvVarTikz}[1]{\texttt{\itshape{\textcolor{BehaviorTreeEnvironmentColor}{#1}}}}
\newcommand{\BlVarTikz}[1]{\texttt{\itshape{\textcolor{BehaviorTreeBlackboardColor}{#1}}}}

\newcommand{\NodeStatusTikz}{\texttt{\itshape{status}}}

\newcommand{\StatefulBehaviorTree}{\mathit{SBT}}

\newcommand{\NeuroSymbolicBT}{NSBT}

\newcommand{\BehaviorTreeMonitor}{\mathit{BTM}}

\usetikzlibrary{shapes.geometric, shapes.misc, shapes.arrows, shapes.callouts, shapes.multipart, fit, positioning, calc, matrix, automata, arrows.meta}

\definecolor{BehaviorTreeSelectorColor}{HTML}{00FFFF}
\definecolor{BehaviorTreeSequenceColor}{HTML}{FFA500}
\definecolor{BehaviorTreeParallelColor}{HTML}{FFD700}
\definecolor{BehaviorTreeLeafColor}{HTML}{C0C0C0}

\tikzset{
    Selector/.style={
        draw=black,
        fill=BehaviorTreeSelectorColor,
        text=black,
        shape=chamfered rectangle,
        minimum size=15pt,
        inner sep=0pt,
        font=\tiny
    },
    Sequence/.style={
        draw=black,
        fill=BehaviorTreeSequenceColor,
        text=black,
        shape=rectangle,
        minimum size=15pt,
        inner sep=0pt,
        inner xsep=2pt,
        font=\tiny
    },
    Parallel/.style={
        draw=black,
        fill=BehaviorTreeParallelColor,
        text=black,
        shape=trapezium,
        trapezium left angle=60,
        trapezium right angle=120,
        minimum size=15pt,
        inner sep=0pt,
        trapezium stretches body,
        font=\tiny,
        align=center
    },
    Decorator/.style={
        draw=black,
        fill=white,
        text=black,
        shape=trapezium,
        trapezium left angle=67,
        trapezium right angle=67,
        minimum size=15pt,
        inner sep=0pt,
        trapezium stretches body,
        font=\tiny,
        align=center
    },
    Action/.style={
        draw=black,
        fill=BehaviorTreeLeafColor,
        text=black,
        shape=ellipse,
        minimum size=15pt,
        inner sep=0pt,
        font=\tiny
    },
    Check/.style={
        draw=black,
        fill=BehaviorTreeLeafColor,
        text=red,
        shape=ellipse,
        minimum size=15pt,
        inner sep=0pt,
        font=\tiny
    },
    Blackboard/.style={
        draw=blue,
        fill=white,
        text=black,
        shape=rectangle,
        font=\tiny,
        inner sep = .5pt,
        inner xsep = 0pt,
        anchor = north
    },
  TreeTable/.style={
    matrix of nodes,
    row sep=0pt,
    column sep=0pt,
    nodes={
      rectangle,
      draw=black,
      align=center
    },
    minimum height=1pt,
    text depth=0.5ex,
    text height=1.5ex,
    nodes in empty cells,
    column 1/.style={
      nodes={text width=5.0em,font=\bfseries}
    }
  },
  TreeTableWide/.style={
    matrix of nodes,
    row sep=0pt,
    column sep=0pt,
    nodes={
      rectangle,
      draw=black,
      align=center
    },
    minimum width=7pt,
    minimum height=1pt,
    text depth=0.5ex,
    text height=1.5ex,
    nodes in empty cells,
    column 1/.style={
      nodes={text width=5.0em,font=\bfseries}
    }
  }
}

\newcommand{\True}{\top}
\newcommand{\False}{\bot}

\lstdefinestyle{Terminal}{
    basicstyle=\tiny\ttfamily,
    showstringspaces=false,
    breaklines=true,
    columns=fullflexible,
    backgroundcolor=\color{lightgray!20},
    frame=single
}

\definecolor{BehaVerifyColor1}{HTML}{D2691E} 
\definecolor{BehaVerifyColor2}{HTML}{0000CD} 
\definecolor{BehaVerifyColor3}{HTML}{FF1493} 
\definecolor{BehaVerifyColor4}{HTML}{FF1493} 
\definecolor{BehaVerifyColor5}{HTML}{FF4500} 
\definecolor{BehaVerifyColor6}{HTML}{FF1493} 
\definecolor{BehaVerifyColor7}{HTML}{FF1493} 
\definecolor{BehaVerifyColor8}{HTML}{778899} 

\definecolor{BehaVerifyColorBlack}{RGB}{0, 0, 0}
\definecolor{BehaVerifyColorWhite}{RGB}{255, 255, 255}

\lstdefinelanguage{BehaVerify}{
    keywordstyle=[1]\color{BehaVerifyColor1},
    keywordstyle=[2]\color{BehaVerifyColor2},
    keywordstyle=[3]\color{BehaVerifyColor3},
    keywordstyle=[4]\color{BehaVerifyColor4},
    keywordstyle=[5]\color{BehaVerifyColor5},
    keywordstyle=[6]\color{BehaVerifyColor6},
    keywordstyle=[7]\color{BehaVerifyColor7},
    keywords=[1]{True, False, success, running, failure},
    keywords=[2]{abs, max, min, sin, cos, exp, tan, ln, eq, neq, lt, gt, lte, gte, neg, add, sub, mult, idiv, rdiv, mod, count, index, not, and, or, xor, xnor, implies, equiv, active, success, running, failure, next, globally, globally_bounded, finally, finally_bounded, until, until_bounded, release, release_bounded, previous, not_previous_not, historically, historically_bounded, once, once_bounded, since, since_bounded, triggered, triggered_bounded, exists_globally, exists_next, exists_finally, exists_until, always_globally, always_next, always_finally, always_until},
    keywords=[3]{assign, end_assign, case, end_case, result, end_result, condition, end_condition, LTLSPEC, end_LTLSPEC, CTLSPEC, end_CTLSPEC, INVARSPEC, end_INVARSPEC},
    keywords=[4]{variable, end_variable, variable_statement, end_variable_statement, initial_values, end_initial_values, update, end_update, write_environment, end_write_environment, read_environment, end_read_environment, return_statement, end_return_statement},
    keywords=[5]{parallel, selector, sequence, X_is_Y, inverter, read_variables, end_read_variables, write_variables, end_write_variables, arguments, end_arguments, local_variables, end_local_variables, policy, success_on_all, success_on_one, with_partial_memory, with_true_memory, instant, constant_index, range, children, end_children, child, end_child, X, Y, bl, env, local, VAR, BOOLEAN},
    keywords=[6]{composite, end_composite, decorator, end_decorator, check, end_check, action, end_action, environment_check, end_environment_check, sub_tree, end_sub_tree, insert, end_insert},
    keywords=[7]{configuration, enumerations, constants, end_constants, variables, end_variables, environment_update, end_environment_update, checks, end_checks, environment_checks, end_environment_checks, actions, end_actions, sub_trees, end_sub_trees, tree, end_tree, tick_prerequisite, end_tick_prerequisite, specifications, end_specifications},
  sensitive=true,
  comment=[s]{\#\{}{\}\#},
  commentstyle=\color{BehaVerifyColor8}\slshape,
  stringstyle=\color{BehaVerifyColor1},
  morestring=[b]'
}

\lstdefinestyle{BehaVerifyStyle}{
  basicstyle=\tiny\color{BehaVerifyColorBlack}\ttfamily,
  backgroundcolor=\color{BehaVerifyColorWhite},
  language=BehaVerify,
  numbers=left,
  xleftmargin=10pt,
  numberstyle=\tiny\color{BehaVerifyColorBlack},
  stepnumber=1,
  numbersep=10pt,
  showstringspaces=false,
  breaklines=true,
  columns=fullflexible,
  frame=single,
  framexleftmargin=1pt
}

\definecolor{nuXmvColor1}{RGB}{0,0,255}
\definecolor{nuXmvColor2}{RGB}{255,51,51}
\definecolor{nuXmvColor3}{RGB}{0,204,102}
\definecolor{nuXmvColor4}{RGB}{153,153,0}
\definecolor{nuXmvColor5}{RGB}{255,153,204}
\definecolor{nuXmvColor6}{RGB}{0,102,51}
\definecolor{nuXmvColor7}{RGB}{0,200,200}
\definecolor{nuXmvColor8}{RGB}{255,255,0}
\definecolor{nuXmvColor9}{RGB}{169,3,252}

\definecolor{nuXmvColorBlack}{RGB}{0, 0, 0}
\definecolor{nuXmvColorWhite}{RGB}{255, 255, 255}

\lstdefinelanguage{nuXmv}{
    keywordstyle=[1]\color{nuXmvColor1},
    keywordstyle=[2]\color{nuXmvColor2},
    keywordstyle=[3]\color{nuXmvColor3},
    keywordstyle=[4]\color{nuXmvColor4},
    keywordstyle=[5]\color{nuXmvColor5},
    keywordstyle=[6]\color{nuXmvColor6},
    keywordstyle=[7]\color{nuXmvColor7},
    keywords=[1]{TRUE, FALSE},
    keywords=[2]{pi, abs, max, min, sin, cos, exp, tan, ln, xor, xnor, mod,  word1, bool, toint, count, swcost, uwconst, signed, unsigned, sizeof, floor, extend, resize, union, in, READ, WRITE, CONSTARRAY, typeof, EG, EX, EF, AG, AX, AF, U, X, G, F, V, Y, Z, H, O, S, T},
    keywords=[3]{ASSIGN, TRANS, INIT, FAIRNESS, JUSTICE, COMPASSION, INVAR, next, init, case, esac},
    keywords=[4]{CONSTANTS, DEFINE, VAR, IVAR, FROZENVAR, FUN},
    keywords=[5]{integer, real, boolean, array, unsigned word, signed word, clock},
    keywords=[6]{LTLSPEC, CTLSPEC, SPEC, INVARSPEC},
    keywords=[7]{MODULE},
  sensitive=true,
  comment=[l]{--},
  commentstyle=\color{nuXmvColor9}
}

\lstdefinestyle{nuXmvStyle}{
  basicstyle=\tiny\color{nuXmvColorBlack}\ttfamily,
  backgroundcolor=\color{nuXmvColorWhite},
  language=nuXmv,
  numbers=left,
  xleftmargin=10pt,
  numberstyle=\tiny\color{nuXmvColorBlack},
  stepnumber=1,
  numbersep=10pt,
  tabsize=2,
  showstringspaces=false,
  breaklines=true,
  columns=fullflexible,
  frame=single,
  framexleftmargin=1pt
}

\definecolor{textXColor1}{HTML}{D2691E} 
\definecolor{textXColor2}{HTML}{0000CD} 
\definecolor{textXColor3}{HTML}{FF1493} 
\definecolor{textXColor4}{HTML}{FF1493} 
\definecolor{textXColor5}{HTML}{FF4500} 
\definecolor{textXColor6}{HTML}{FF1493} 
\definecolor{textXColor7}{HTML}{FF1493} 
\definecolor{textXColor8}{HTML}{778899} 

\definecolor{textXColorBlack}{RGB}{0, 0, 0}
\definecolor{textXColorWhite}{RGB}{255, 255, 255}

\lstdefinelanguage{textX}{
    keywordstyle=[1]\color{textXColor6},
    keywordstyle=[2]\color{textXColor8},
  keywords=[1]{STRICTFLOAT, FLOAT, INT, STRING, BOOL},
  sensitive=true,
  comment=[l]{//},
  commentstyle=\color{textXColor8}\slshape,
  stringstyle=\color{textXColor2},
  morestring=[b]'
}

\lstdefinestyle{textXStyle}{
  basicstyle=\tiny\color{textXColorBlack}\ttfamily,
  backgroundcolor=\color{textXColorWhite},
  language=textX,
  numbers=left,
  numberstyle=\tiny\color{textXColorBlack},
  stepnumber=1,
  numbersep=10pt,
  tabsize=2
}

\newcommand{\ChapterTitle}{\toolname: A Formal Verification Tool for Behavior Trees}

\newcommand{\toolname}[0]{\textsf{BehaVerify}\xspace}

\title{\ChapterTitle}

\author{
Serena S. Serbinowska \quad Ben Wooding \quad Diego Manzanas Lopez \quad Luke Nam
\\
Anne M. Tumlin \quad Nicholas K. Potteiger \quad Hongchao Zhang \quad Taylor T. Johnson
\institute{Institute for Convergent Software Integrated Systems, Vanderbilt University\\
Nashville, TN 37235, USA}
\email{\{serena.serbinowska, ben.wooding, diego.manzanas.lopez, luke.nam,}
\email{anne.m.tumlin, nicholas.potteiger, hongchao.zhang, taylor.johnson\}@vanderbilt.edu}
}

\def\titlerunning{BehaVerify: A Formal Verification Toolbox for Behavior Trees}
\def\authorrunning{S.S. Serbinowska et al.}

\begin{document}

\maketitle

\begin{abstract}
 This paper presents \textsf{BehaVerify}, a verification toolbox for \emph{behavior trees} (BTs) and Neuro-Symbolic BTs. BTs are a graphical language for planning and sequencing behaviors --- similar to extended state machines --- and are increasingly used in safety-critical domains. \textsf{BehaVerify} enables users to specify a BT using a \emph{Domain Specific Language} (DSL) with formal syntax and semantics. This language further includes \emph{Linear Temporal Logic} and \emph{Computation Tree Logic} specifications. \textsf{BehaVerify} uses an efficient encoding into \textsf{nuXmv}, called \emph{fastforwarding}, to quickly verify if a specification holds over the BT. Additionally, \textsf{BehaVerify} can generate implementations of the BT in Python and Haskell. Through Python, \toolname~has been integrated with PyTrees and Robot Operating System (ROS) and has been deployed to several autonomous robotic systems, including for verifying runtime monitors. While prior works have developed the underlying theory, this manuscript focuses on the toolbox usage, functionality and implementation. \toolname is open source:
\end{abstract}

\centerline{\url{https://github.com/verivital/behaverify}}


\section{Introduction}
\emph{Behavior Trees} ($\BehaviorTree$s) are high-level controllers that have gained popularity and are increasingly used for safety-critical tasks\@.  Accordingly, it is imperative to verify their correctness\@.
In medicine, $\BehaviorTree$s are used for sensitive machine
insertion tasks~\cite{wu2024ICRA} and to aid with hand
rehabilitation~\cite{zubrycki2024RoMoCo}\@. In
security,~\cite{samaddar2025ICAIC,potteiger2024ICAA} use $\BehaviorTree$s with
learning-enabled components for autonomous computer network defense\@. Sensitive
drone tasks rely on $\BehaviorTree$s for drones that inspect
structures~\cite{rocamora2024ICUAS} and in mining a BT controls robot that assembles explosive charges and places them in the rock face~\cite{hallen2024CASE}\@.
Example use cases for $\BehaviorTree$s are provided in detail in the surveys \cite{shin2024ICTC,iovino2022Robotics-and-Autonomous-Systems}.
We distinguish our work from the formalism `Behavior Trees' that
refers to a graphical modeling language used for requirement
handling~\cite{grunske2008Journal-of-Visual-Language-and-Computing}\@.
Intuitively, a $\BehaviorTree$ is a rooted tree re-evaluated from the
root on every `tick'. Interior nodes are control-flow operators. A
\emph{selector} tries its children in order until one succeeds, while a
\emph{sequence} runs its children in order until one fails. In contrast, leaf nodes
check a condition or perform an action. Every node returns \emph{Success},
\emph{Failure} or \emph{Running}, and that status propagates back up the tree to
determine what runs next. The result offers the modularity and readability of a
state machine without the quadratic growth in explicit transitions, which is
partly why $\BehaviorTree$s have displaced hand-written state machines across
robotics.

In this work, we present \toolname, which builds upon the theoretical foundations~\cite{serbinowska2022SEFM,serbinowska2025NEUS-REPLACE,serbinowska2024FMAS-BTM,serbinowska2024FMAS-SBT}. To maintain brevity and avoid repetition of theoretical details, this work focuses on the usage, functionality and implementation of \toolname.

\paragraph{Related Verification Tools.}
Several existing tools and frameworks verify $\BehaviorTree$s\@.
\textsf{ArcadeBT}~\cite{henn2022FMICS} is a verification tool for $\BehaviorTree$s that uses linear \emph{Constrained Horn Clauses} (CHCs)\@.
\textsf{BT2BIP}~\cite{wang2024FACS}, or \textsf{xml2bip}, transforms $\BehaviorTree$s into the \emph{Behavior-Interaction-Priority} (BIP) framework and then treats them as an extended finite state machine to enable verification using \textsf{nuXmv}~\cite{nuXmv}\@.
\textsf{BT2Fiacre}~\cite{ingrand2025arXiv} translates $\BehaviorTree$s into \textsf{Fiacre}~\cite{berthomieu2008ERTS2} which enables verification\@.
\textsf{BTCompiler}~\cite{btcompiler2019} utilizes OCaml and ROCQ (formerly COQ)~\cite{Coq}\@.
\textsf{MoVe4BT}~\cite{MoVe4BT} uses the Timed CSPbt modeling and allows for LTL specifications and true parallel execution of nodes\@.
We recreate examples from \textsf{BT2BIP} and \textsf{BT2Fiacre}, and compare \textsf{BehaVerify} against \textsf{MoVe4BT} and \textsf{BT2Fiacre}, while \cite{serbinowska2022SEFM} sufficiently compares \textsf{BTCompiler}\@. A framework for formal verification of \emph{Linear Temporal Logic} (LTL) specifications for $\BehaviorTree$s is included in~\cite{biggar2020IEEE-RAL}; unlike \textsf{BehaVerify}, it only support Booleans\@.

While we are primarily focused on design time/offline verification, there exist other tools for $\BehaviorTree$s\@.
A formal property specification language was introduced in~\cite{colledanchise2021IROS}; it can be used to build runtime monitors for $\BehaviorTree$\@.
$\BehaviorTree$ synthesis tools \cite{scheide2025AutonomousRobots,hong2023} create $\BehaviorTree$s to meet specifications\@.

\paragraph{This Work versus Prior \toolname Papers.}
\toolname has never itself been the subject of a tool paper. Our earlier
publications~\cite{serbinowska2022SEFM,serbinowska2024FMAS-SBT,serbinowska2024FMAS-BTM,serbinowska2025NEUS-REPLACE}
are theory papers, albeit the appearance of the tool's name in the title
of~\cite{serbinowska2022SEFM} invites confusion on this point. They contribute
the semantics of Stateful Behavior Trees ($\StatefulBehaviorTree$s), their neuro-symbolic extension, and
the runtime-monitor construction. Each paper was accompanied by a separate research
prototype, and none evaluated \toolname as a tool. We summarize and cite that
theory here rather than re-develop it. What this paper contributes is the tool
itself: one distribution-ready toolbox that implements all three lines of that
theory behind a single Domain Specific Language (DSL) and command-line interface (CLI), the engineering that makes it usable and
reusable by others, and an evaluation the theory papers did not perform.

\paragraph{Contributions.} \textsf{BehaVerify}\footnote{\url{https://github.com/verivital/behaverify}}
formalizes $\BehaviorTree$s and provides a framework to verify them with
\textsf{nuXmv}, as well as run them in Python and Haskell\@. The following three
capabilities implement previously published semantics.
What is new is their \emph{consolidation}: all three now live in one toolbox,
behind a single DSL and CLI, rather than in three separate prototypes.
\begin{itemize}
    \item \textbf{DSL for SBTs }($\StatefulBehaviorTree$s): \textsf{BehaVerify} can verify the $\StatefulBehaviorTree$s formalized in~\cite{serbinowska2024FMAS-SBT}, implemented as a DSL for $\BehaviorTree$s and their semantics, including a restricted grammar for writing updates to the state variables (the `blackboard');
    \item \textbf{Neuro-Symbolic Behavior Trees} ($\NeuroSymbolicBT$s): \textsf{BehaVerify} handles $\BehaviorTree$s that make use of neural networks within leaf nodes, as introduced in~\cite{serbinowska2025NEUS-REPLACE}, with extensions of the DSL and semantics as $\StatefulBehaviorTree$s;
    \item \textbf{Verified Runtime Monitors}: \textsf{BehaVerify} has the capability to generate and verify runtime monitors, leveraging the method developed in~\cite{serbinowska2024FMAS-BTM}\@.
\end{itemize}
The contributions that are new to this paper, and that did not exist at the time
of the prior papers, are the following.
\begin{itemize}
    \item \textbf{First distribution-ready release:} \textsf{BehaVerify} can be installed via PyPI\footnote{\url{https://pypi.org/project/behaverify/}} (\texttt{pip install \\behaverify}) or Docker and run directly from the command line with multiple modes, as described in the Documentation webpages\footnote{\url{https://verivital.github.io/behaverify/}}; a tutorial with examples is also available\footnote{\url{https://github.com/verivital/behaverify/tree/main/tutorial_examples}}\@. A graphical user interface (GUI) visualizes trees and counterexample traces. Continuous Integration and Continuous Delivery/Deployment (CI/CD) via GitHub Actions, with unit and end-to-end tests, enables frequent and reliable code changes and updates.\restored{\footnote{A key contribution of \toolname is its reusability, and a reproducibility package accompanies every result reported here.}}
    \item \textbf{Scalability, and the first controlled ablation of fastforwarding:} a naive-encoding mode isolates the effect of the encoding itself (Table~\ref{tab:ff-ablation}). The comparison in \cite{serbinowska2024FMAS-SBT} was against \textsf{MoVe4BT}, a different tool with different semantics, so it cannot separate the effect of the encoding from the effect of the tool. Running both encodings inside \toolname separates the two, adds state-space measurements, and quantifies the scalability advantage of \toolname over the prior theoretical implementations.
    \item \textbf{New comparisons:} the \textsf{BT2Fiacre} drone recreation (Table~\ref{tab:drone3compare}), including a previously unreported battery-priority violation, and recreations of \textsf{BT2BIP}'s Mars Rover and Train Control examples. Neither tool appears in any prior \toolname paper. Of the tool comparisons, only the \textsf{MoVe4BT} summary is carried over. The \textsf{MoVe4BT} comparison scales to 2\,055 nodes, while the \textsf{BT2Fiacre} comparison involves a tree of more than 30 nodes operating in a three-dimensional environment\@.
\end{itemize}
\section{Behavior Tree Overview}
\label{sec:bt-overview}

Figure~\ref{fig:drunk-drone}  and Table~\ref{fig:drunk-drone-table} provide a $\BehaviorTree$ sample execution of a `Drunken Drone'. A $\BehaviorTree$ is a rooted tree that executes when it receives an exterior signal: a `tick.'\@
We use discrete time steps to describe stepping through the tree\@.
At each time step during a tick, exactly one node in the tree is active\@.
If a node is active at a given time step, the node that will be active at the next time step is either a parent or a child of said node\@ (with exception of the root node).

\begin{figure}[t]
  \centering
  \begin{adjustbox}{max width = \linewidth}
    \begin{tikzpicture}
    \tikzset{level distance=20pt}
    \tikzset{sibling distance=0.5pt}
    \Tree[.\node[Selector](Drone){$Drone$};
    [.\node[Sequence](GoalSeq){$GoalSeq$};
    [.\node[Check](NeedNew){\begin{tabular}{c}$NeedNew$\end{tabular}};\node[Blackboard](NeedNewUPDATEBLACKBOARD){\begin{tabular}{l}\texttt{if }\begin{math}((\EnvVarTikz{x\_g} = \EnvVarTikz{x\_d}) \land (\EnvVarTikz{y\_g} = \EnvVarTikz{y\_d}))\end{math}\texttt{:}\\{}\begin{math}\quad{}\NodeStatusTikz{}\coloneqq{} \SuccessTikz{}\end{math}\\{}\texttt{else:}\\{}\begin{math}\quad{}\NodeStatusTikz{}\coloneqq{}\FailureTikz{}\end{math}\end{tabular}};]
    [.\node[Action](NewGoal){\begin{tabular}{c}$NewGoal$\end{tabular}};\node[Blackboard](NewGoalUPDATEBLACKBOARD){\begin{tabular}{l}\begin{math}\end{math}\begin{math}\BlVarTikz{new} \in \end{math}\begin{math}\{\True{}\}\end{math}\\\begin{math}\end{math}\begin{math}\BlVarTikz{act} \in \end{math}\begin{math}\{\texttt{\textquotesingle{}XX\textquotesingle{}}\}\end{math}\\\begin{math}\end{math}\begin{math}\NodeStatusTikz{} \coloneqq{} \RunningTikz{}\end{math}\end{tabular}};]
    ]
    [.\node[Action](NextAct){\begin{tabular}{c}$NextAct$\end{tabular}};\node[Blackboard](NextActUPDATEBLACKBOARD){\begin{tabular}{l}\begin{math}\end{math}\begin{math}\BlVarTikz{new} \in \end{math}\begin{math}\{\False{}\}\end{math}\\\begin{math}\end{math}\begin{math}\BlVarTikz{act} \in \end{math}\begin{math}\{\texttt{\textquotesingle{}No\textquotesingle{}}, \texttt{\textquotesingle{}So\textquotesingle{}}, \texttt{\textquotesingle{}We\textquotesingle{}}, \texttt{\textquotesingle{}Ea\textquotesingle{}}\}\end{math}\\\begin{math}\end{math}\begin{math}\NodeStatusTikz{} \coloneqq{} \SuccessTikz{}\end{math}\end{tabular}};]
    ]
    \node[draw=none,inner sep=0pt, fit=(NeedNewUPDATEBLACKBOARD)(NewGoalUPDATEBLACKBOARD)(NextActUPDATEBLACKBOARD)] (sereneblackboardfitbox) {};
    \node[Blackboard, anchor=south east] (initialValues) at ([xshift=-5pt]sereneblackboardfitbox.south west){\begin{tabular}{l}\texttt{-INITIAL VALUES-}\\{}\begin{math}\end{math}\begin{math}\BlVarTikz{act} \in \end{math}\begin{math}\{\texttt{\textquotesingle{}XX\textquotesingle{}}\}\end{math}\\\begin{math}\end{math}\begin{math}\BlVarTikz{new} \in \end{math}\begin{math}\{\False{}\}\end{math}\\\begin{math}\end{math}\begin{math}\EnvVarTikz{x\_d} \in \end{math}\begin{math}\{0\}\end{math}\\\begin{math}\end{math}\begin{math}\EnvVarTikz{y\_d} \in \end{math}\begin{math}\{0\}\end{math}\\\begin{math}\end{math}\begin{math}\EnvVarTikz{x\_g} \in \end{math}\begin{math}\{\EnvVarTikz{x\_d}\}\end{math}\\\begin{math}\end{math}\begin{math}\EnvVarTikz{y\_g} \in \end{math}\begin{math}\{\EnvVarTikz{y\_d}\}\end{math}\\\end{tabular}};
    \node[Blackboard, anchor=south east] (varInfo) at ([xshift=-5pt]initialValues.south west){\begin{tabular}{l}\texttt{-VARIABLE INFO-}\\{}\BlVarTikz{act} - \texttt{VAR} - \texttt{\{}\texttt{\textquotesingle{}We\textquotesingle{}}, \texttt{\textquotesingle{}Ea\textquotesingle{}}, \texttt{\textquotesingle{}No\textquotesingle{}}, \texttt{\textquotesingle{}So\textquotesingle{}}, \texttt{\textquotesingle{}XX\textquotesingle{}}\texttt{\}}\\{}\BlVarTikz{new} - \texttt{VAR} - \texttt{BOOLEAN}\\{}\EnvVarTikz{x\_d} - \texttt{VAR} - [0, 19]\\{}\EnvVarTikz{y\_d} - \texttt{VAR} - [0, 19]\\{}\EnvVarTikz{x\_g} - \texttt{VAR} - [0, 19]\\{}\EnvVarTikz{y\_g} - \texttt{VAR} - [0, 19]\end{tabular}};
    \node[Blackboard, anchor=south west] (environmentUpdates) at ([xshift=5pt]sereneblackboardfitbox.south east){\begin{tabular}{l}\texttt{-ENVIRONMENT UPDATES-}\\{}\begin{math}\end{math}\begin{math}\EnvVarTikz{x\_g} \in \end{math}\texttt{(}\begin{math}[0, 19]\end{math}\\\begin{math}\quad{}\end{math}\texttt{if }\begin{math}\BlVarTikz{new}\end{math}\texttt{ else}\\\begin{math}\quad{}\end{math}\begin{math}\{\EnvVarTikz{x\_g}\}\end{math}\texttt{)}\\\begin{math}\end{math}\begin{math}\EnvVarTikz{y\_g} \in \end{math}\texttt{(}\begin{math}[0, 19]\end{math}\\\begin{math}\quad{}\end{math}\texttt{if }\begin{math}\BlVarTikz{new}\end{math}\texttt{ else}\\\begin{math}\quad{}\end{math}\begin{math}\{\EnvVarTikz{y\_g}\}\end{math}\texttt{)}\\\begin{math}\end{math}\begin{math}\EnvVarTikz{x\_d} \in \end{math}\texttt{(}\begin{math}\{max(0, (\EnvVarTikz{x\_d} - 1))\}\end{math}\\\begin{math}\quad{}\end{math}\texttt{if }\begin{math}(\BlVarTikz{act} = \texttt{\textquotesingle{}We\textquotesingle{}})\end{math}\texttt{ else}\\\begin{math}\quad{}\quad{}\end{math}\texttt{(}\begin{math}\{min(19, (\EnvVarTikz{x\_d} + 1))\}\end{math}\\\begin{math}\quad{}\quad{}\end{math}\texttt{if }\begin{math}(\BlVarTikz{act} = \texttt{\textquotesingle{}Ea\textquotesingle{}})\end{math}\texttt{ else}\\\begin{math}\quad{}\quad{}\end{math}\begin{math}\{\EnvVarTikz{x\_d}\}\end{math}\texttt{))}\\\begin{math}\end{math}\begin{math}\EnvVarTikz{y\_d} \in \end{math}\texttt{(}\begin{math}\{max(0, (\EnvVarTikz{y\_d} - 1))\}\end{math}\\\begin{math}\quad{}\end{math}\texttt{if }\begin{math}(\BlVarTikz{act} = \texttt{\textquotesingle{}So\textquotesingle{}})\end{math}\texttt{ else}\\\begin{math}\quad{}\quad{}\end{math}\texttt{(}\begin{math}\{min(19, (\EnvVarTikz{y\_d} + 1))\}\end{math}\\\begin{math}\quad{}\quad{}\end{math}\texttt{if }\begin{math}(\BlVarTikz{act} = \texttt{\textquotesingle{}No\textquotesingle{}})\end{math}\texttt{ else}\\\begin{math}\quad{}\quad{}\end{math}\begin{math}\{\EnvVarTikz{y\_d}\}\end{math}\texttt{))}\\\end{tabular}};
    \end{tikzpicture}
  \end{adjustbox}
  \caption{
    \emph{A simple drone controller $\BehaviorTree$}\@.
    Node $NeedNew$ checks if the drone's location $(\EnvVar{x\_d}, \EnvVar{y\_d})$ is equal to the location of the goal $(\EnvVar{x\_g}, \EnvVar{y\_g})$\@.
    If it is, a new goal is needed\@.
    Otherwise, an action is nondeterministically chosen\@.
    Between ticks, the environment updates occur; the location of the drone is updated based on the chosen direction and a new goal is generated if needed\@.
    An example execution can be found in Table~\ref{fig:drunk-drone-table}\@.
    \texttt{XX} is a
    placeholder meaning \emph{no action was chosen} on that time step. ``Sampling'' from an interval denotes a nondeterministic
    choice over that interval, which the model checker explores exhaustively
    rather than treating as a random draw.
  }\label{fig:drunk-drone}
\end{figure}
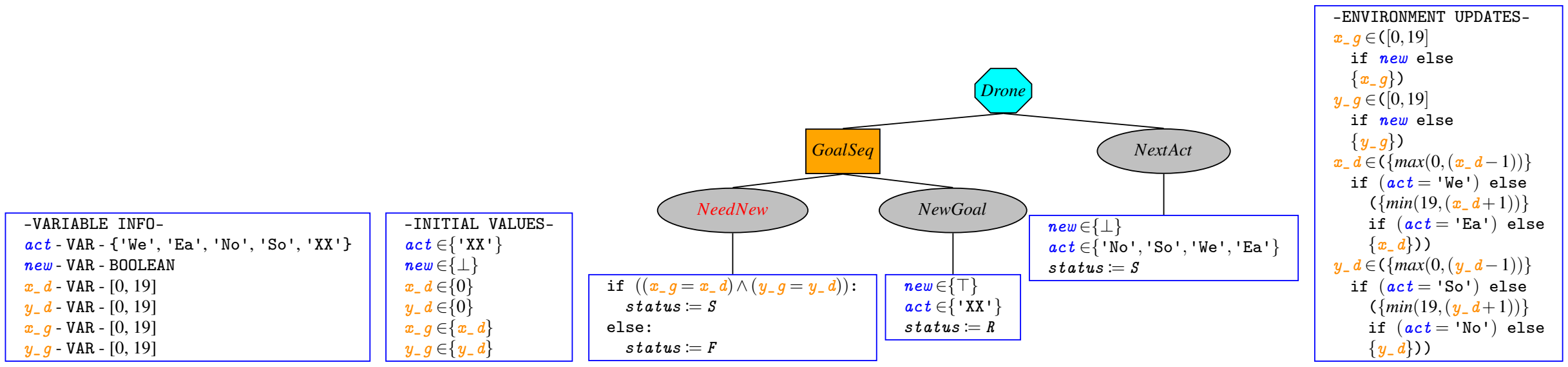
\begin{table}[t]
  \centering
  {\small
  \begin{tabular*}{\linewidth}{@{\extracolsep{\fill}}l| *{16}{c}}
    \toprule
    \textbf{Time}    & 1 & 2 & 3 & 4 & 5 & 6 & 7 & 8 & 9 & 10 & 11 & 12 & 13 & 14 & 15 & 16\\
    \textbf{Tick} & 1 & 1 & 1 & 1 & 1 & 1 & 1 & 1 & 2 & 2 &  2 &  2 &  2 &  2 &  2 &  2\\
    \textbf{Active} & D & GS & NN & GS & NG & GS & D & Env & D & GS & NN & GS & D & NA & D & Env\\
    \textbf{Return} & $\Invalid$ & $\Invalid$ & $\Success$ & $\Invalid$ & $\Running$ & $\Running$ & $\Running$ & $\Invalid$ & $\Invalid$ & $\Invalid$ & $\Failure$ & $\Failure$ & $\Invalid$ & $\Success$ & $\Success$ & $\Invalid$\\
    \textbf{$\BlVar{act}$} & \texttt{XX} & \texttt{XX} & \texttt{XX} & \texttt{XX} & \texttt{XX} & \texttt{XX} & \texttt{XX} & \texttt{XX} & \texttt{XX} & \texttt{XX} & \texttt{XX} & \texttt{XX} & \texttt{XX} & \texttt{No} & \texttt{No} & \texttt{No}\\
    \textbf{$\BlVar{new}$} & $\False{}$ & $\False{}$ & $\False{}$ & $\False{}$ & $\True{}$ & $\True{}$ & $\True{}$ & $\True{}$ & $\True{}$ & $\True{}$ & $\True{}$ & $\True{}$ & $\True{}$ & $\False{}$ & $\False{}$ & $\False{}$\\
    \textbf{$\EnvVar{x\_d}$} & 0 & 0 & 0 & 0 & 0 & 0 & 0 & 0 & 0 & 0 & 0 & 0 & 0 & 0 & 0 & 0\\
    \textbf{$\EnvVar{y\_d}$} & 0 & 0 & 0 & 0 & 0 & 0 & 0 & 0 & 0 & 0 & 0 & 0 & 0 & 0 & 0 & 1\\
    \textbf{$\EnvVar{x\_g}$} & 0 & 0 & 0 & 0 & 0 & 0 & 0 & 2 & 2 & 2 & 2 & 2 & 2 & 2 & 2 & 2\\
    \textbf{$\EnvVar{y\_g}$} & 0 & 0 & 0 & 0 & 0 & 0 & 0 & 3 & 3 & 3 & 3 & 3 & 3 & 3 & 3 & 3\\
    \midrule
    \textbf{$\textbf{FF}$} & $s_0$ &  &  &  &  &  &  &$s_1$ &  &  &  &  &  &  &  & $s_2$\\
    \bottomrule
  \end{tabular*}
  }
  \caption{
    \emph{A table illustrating a hypothetical execution of the $\BehaviorTree$ shown in} Figure~\ref{fig:drunk-drone}\@.
    \textbf{Time} refers to each time step\@.
    \textbf{Tick} refers to the number of times the tree is called\@.
    \textbf{Active} refers to a single `active' node at each time step\@.
    Node abbreviations: D=Drone (root selector), GS=GoalSeq (sequence), NN=NeedNew (check), NG=NewGoal (action), NA=NextAct (action), Env=environment update (occurs between ticks)\@.
    Nodes return one of $\Success$, $\Failure$, or $\Running$ when they finish; $\Invalid$ indicates the node has not yet returned this tick\@.
    \textbf{Return} shows the current active node status (not a cumulative status)\@. \textbf{FF} marks the only states kept by fastforwarding; each tick is one transition, $s_0\rightarrow s_1\rightarrow s_2$.}
  \label{fig:drunk-drone-table}
\end{table}

Tree traversal is generally depth-first. When a node returns, its status becomes one of \emph{success}, \emph{failure}, or \emph{running} ($\Success$, $\Failure$, or $\Running$)\@.
Until a node returns, its status is \emph{invalid} ($\Invalid$)\@.
The children of a node are ordered; visually we use left to right\@,
\emph{i.e.} the leftmost child is the first child\@. We consider nodes in three broad categories: \emph{leaf}, \emph{composite}, and \emph{decorator}\@.

\paragraph{Variables.}
Variables can be split into two categories: \BlVar{blackboard} and \EnvVar{environment}\@.
The \emph{blackboard} is the shared memory of a tree and can be updated between time steps\@, \emph{e.g.} \BlVar{act}\@.
The \emph{environment} represents variables not under control of the $\BehaviorTree$ and they can be updated between ticks, \emph{e.g.} \EnvVar{x_g}\@.
It is up to the user to decide if a variable should be a blackboard or environment variable\@.

\paragraph{Runtime Assumption.}
As mentioned, there are three return statuses: $\Success$, $\Failure$, and $\Running$\@.
We assume nodes return instantaneously\@.
If a node waits on a long-running calculation, it instantly returns $\Running$ to not block the tree\@.
Moreover, we assume that the environment does not change during the tick, only between ticks\@.

\paragraph{Leaf Nodes.}
Leaf nodes are split into two categories: \emph{checks} and \emph{actions}\@.
\emph{Checks} return $\Success$ if a condition is true and $\Failure$ otherwise\@.
The conditions are predicates depending on blackboard and environment variables\@.
\emph{Actions} can read and set variables and are able to return $\Success$, $\Failure$, or $\Running$\@.
Although it is possible to reduce an entire $\BehaviorTree$ to a single action, key benefits of $\BehaviorTree$s would be lost: \emph{modularity} (ability to reuse subtrees across contexts), \emph{visual clarity} (hierarchical structure providing an intuitive representation of control flow), and \emph{maintainability} (nodes can be modified without affecting the tree significantly)\@.

\paragraph{Composite Nodes.}
The three composite node types are \emph{selector}, \emph{sequence}, and \emph{parallel}\@.
\emph{Selector} nodes activate each of their children in order until a child returns $\Success$ or $\Running$, then the selector immediately returns that same value\@.
If all children return $\Failure$, the selector returns $\Failure$\@.
Thus a selector can return any of $\Success$, $\Failure$ or
$\Running$\@.
\emph{Sequence} nodes work dually, performing identically to Selector nodes but with $\Failure$ and $\Success$ swapped.
A sequence can likewise return any of the three\@.
\emph{Parallel} nodes allow their children to be executed simultaneously. In practice, our tool uses \textsf{PyTrees}~\cite{PyTrees}, which does not have truly parallel execution.\@ Each child is activated in order (no shuffling) and a status is returned based on a user-chosen policy, \emph{e.g. `success on all'}\@.
While selector and sequence nodes can terminate early, skipping children, parallel nodes cannot\@.

\paragraph{Decorator Nodes.}
Decorator nodes have \textbf{exactly} one child\@.
Decorator nodes augment the behavior of the child, \emph{e.g.} negate child, repeat child `n' times, \emph{etc}. They also facilitate node reuse.

\paragraph{Behavior Tree Diagrams.}
A key strength of $\BehaviorTree$s is their visual aspect\@. As such, it is important to be able to visualize $\BehaviorTree$s\@.
\textsf{BehaVerify} allows the user to visualize $\BehaviorTree$s through \texttt{dsl\_to\_latex.py}, which accepts as input a \texttt{.tree} file and produces a \LaTeX\xspace diagram; this was used to generate Figure~\ref{fig:drunk-drone}\@. Each node is represented using a different shape and color\@.
Selector nodes are octagonal, sequence nodes are rectangles, parallel nodes are parallelograms, and
decorator nodes are trapezoids\@.
We represent leaf nodes using ovals; checks have a red label and actions have a black label\@.
Beneath each node is a box containing the code for that leaf node\@.
This code is executed from top to bottom\@. 

\section{BehaVerify Tool Description}
\label{sec:tool-description}
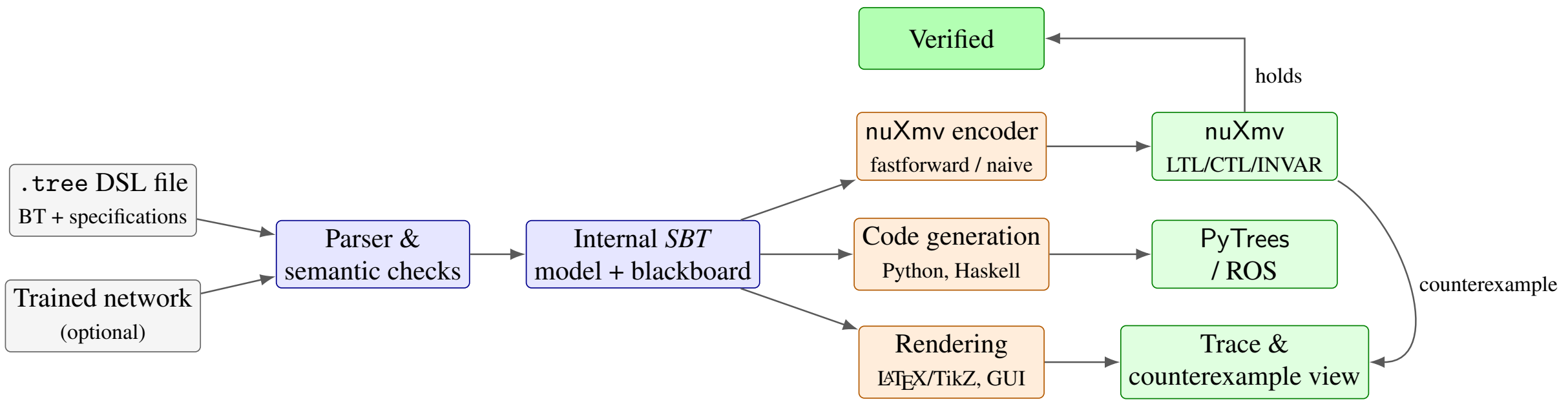
\begin{figure}[t]
\centering
\begin{adjustbox}{max width=\linewidth}
\begin{tikzpicture}[
  font=\small,
  comp/.style={rectangle, rounded corners=2pt, draw=black!60, fill=black!4,
               minimum width=2.4cm, minimum height=0.8cm, align=center,
               inner sep=3pt},
  core/.style={comp, fill=blue!10, draw=blue!55!black},
  back/.style={comp, fill=orange!14, draw=orange!70!black},
  ext/.style={comp, fill=green!12, draw=green!50!black},
  lnk/.style={->, >=Latex, semithick, draw=black!65}
]
  \node[comp] (dsl)   at (0,2.4)   {\texttt{.tree} DSL file\\{\scriptsize BT + specifications}};
  \node[comp] (nn)    at (0,0.9)   {Trained network\\{\scriptsize(optional)}};
  \node[core] (parse) at (3.5,1.7) {Parser \&\\semantic checks};
  \node[core] (model) at (7.0,1.7) {Internal $\StatefulBehaviorTree$\\model + blackboard};

  \node[back] (nuxmv) at (11.0,3.1)  {\textsf{nuXmv} encoder\\{\scriptsize fastforward / naive}};
  \node[back] (code)  at (11.0,1.7)  {Code generation\\{\scriptsize Python, Haskell}};
  \node[back] (draw)  at (11.0,0.3)  {Rendering\\{\scriptsize \LaTeX{}/TikZ, GUI}};

  \node[ext] (solver) at (14.8,3.1)  {\textsf{nuXmv}\\{\scriptsize LTL/CTL/INVAR}};
  \node[ext] (deploy) at (14.8,1.7)  {\textsf{PyTrees}\\/ ROS};
  \node[ext] (trace)  at (14.8,0.3)  {Trace \&\\counterexample view};

  \draw[lnk] (dsl)   -- (parse);
  \draw[lnk] (nn)    -- (parse);
  \draw[lnk] (parse) -- (model);
  \draw[lnk] (model) -- (nuxmv);
  \draw[lnk] (model) -- (code);
  \draw[lnk] (model) -- (draw);
  \draw[lnk] (nuxmv) -- (solver);
  \draw[lnk] (code)  -- (deploy);
  \draw[lnk] (draw)  -- (trace);
  \node[ext, fill=green!30] (ok) at (11.0,4.5) {Verified};
  \draw[lnk] (solver.north) |- node[pos=0.25, right, font=\scriptsize] {holds} (ok.east);
  \draw[lnk] (solver.south east) to[out=-30, in=0, looseness=1.1] node[right, font=\scriptsize] {counterexample} (trace.east);
\end{tikzpicture}
\end{adjustbox}
\caption{Architecture of \toolname. One DSL front end builds a single internal model that every backend reads.}
\label{fig:architecture}
\end{figure}

This section describes \toolname, its inputs from the DSL, output targets, and overall usage. Figure~\ref{fig:architecture} shows how the tool is organized. The user writes the $\BehaviorTree$, its environment and its specifications in one \texttt{.tree} file. A front end parses this file, checks types and variable scopes, and builds one internal model; neural-network leaves enter through the same front end. Every backend reads this model. The \textsf{nuXmv} encoder writes an \texttt{.smv} model, using either the fastforwarding or the naive encoding. The code generators write Python and Haskell implementations, which run the $\BehaviorTree$ on robotic systems through \textsf{PyTrees} and ROS. The renderer draws the tree and any counterexample trace. For each specification, \textsf{nuXmv}~\cite{nuXmv} either confirms that it holds or returns a counterexample, which the trace viewer displays. \toolname is cross-platform, has been tested on Linux, macOS and Windows, installs with \texttt{pip}, and also runs through Docker for reproducibility; the documentation gives detailed instructions.

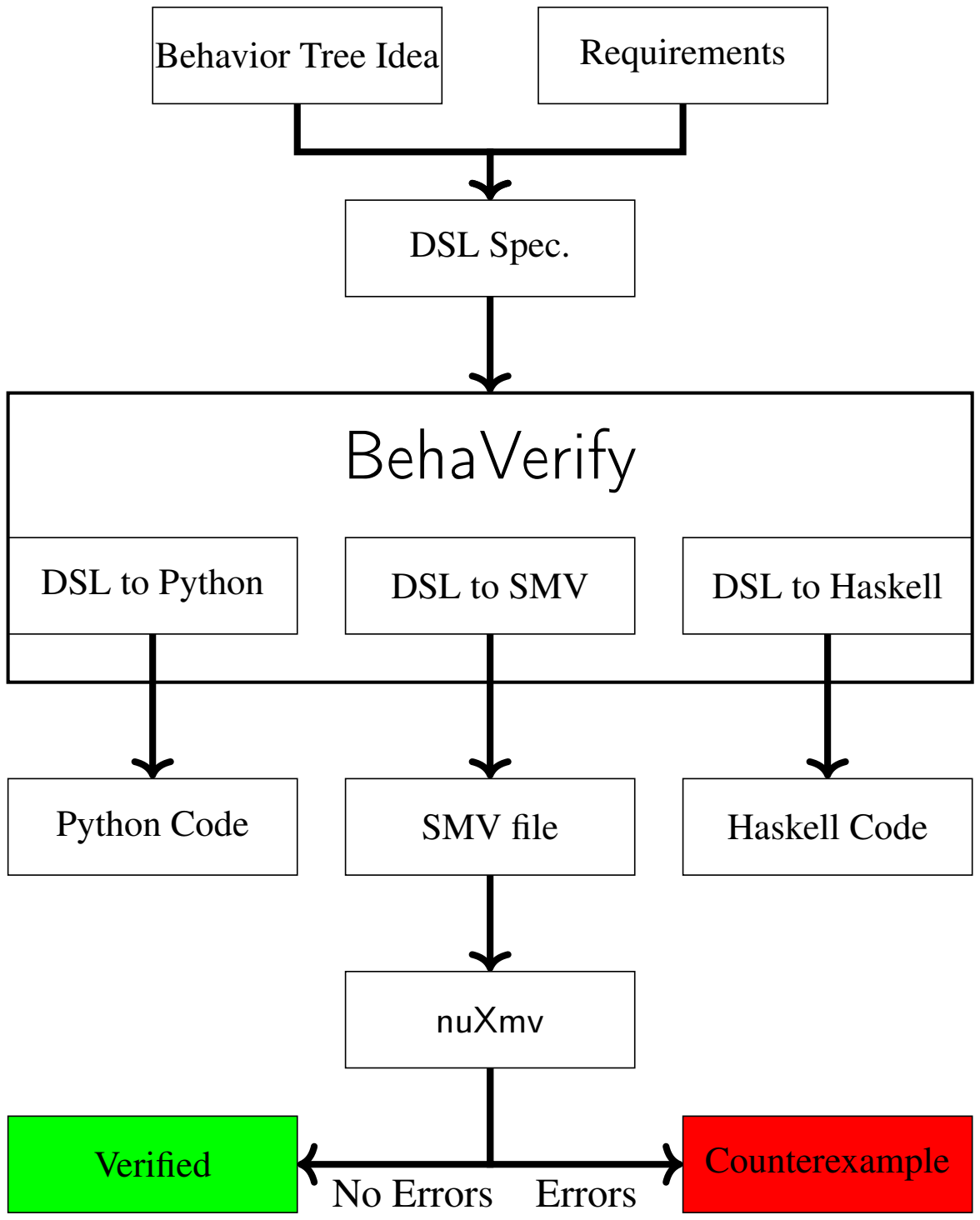
\begin{figure}[t]
\centering
  \begin{minipage}{.48\linewidth}
    \centering
    \begin{adjustbox}{max width = \linewidth}
      \begin{tikzpicture}
      \draw (.5, 8) rectangle ++(3, 1) node[midway] {Behavior Tree Idea};
      \draw[->, line width=2pt] (2, 8) -- (2, 7.5) -- (4, 7.5) -- (4, 7);

      \draw (4.5, 8) rectangle ++(3, 1) node[midway] {Requirements};
      \draw[->, line width=2pt] (6, 8) -- (6, 7.5) -- (4, 7.5) -- (4, 7);

      \draw (2.5, 6) rectangle ++(3, 1) node[midway] {DSL Spec.};
      \draw[->, line width=2pt] (4, 6) -- (4, 5);

      \draw[line width=1pt] (-1, 2) rectangle ++(10, 3) node[midway] {};
      \node[align=center] at (4, 4.3) {\huge \textsf{BehaVerify}};

      \draw (-1, 2.5) rectangle ++(3, 1) node[midway] {DSL to Python};
      \draw[->, line width=2pt] (.5, 2.5) -- (.5, 1);

      \draw (6, 2.5) rectangle ++(3, 1) node[midway] {DSL to Haskell};
      \draw[->, line width=2pt] (7.5, 2.5) -- (7.5, 1);

      \draw (2.5, 2.5) rectangle ++(3, 1) node[midway] {DSL to SMV};
      \draw[->, line width=2pt] (4, 2.5) -- (4, 1);

      \draw (-1, 0) rectangle ++(3, 1) node[midway] {Python Code};

      \draw (6, 0) rectangle ++(3, 1) node[midway] {Haskell Code};

      \draw (2.5, 0) rectangle ++(3, 1) node[midway] {SMV file};
      \draw[->, line width=2pt] (4, 0) -- (4, -1);

      \draw (2.5, -2) rectangle ++(3, 1) node[midway] {\textsf{nuXmv}};
      \draw[->, line width=2pt] (4, -2) -- (4, -3) -- (2, -3);
      \draw[->, line width=2pt] (4, -2) -- (4, -3) -- (6, -3);
      \node at (3.2, -3.3) {\large No Errors};
      \node at (5, -3.3) {\large Errors};

      \draw[fill=green] (-1, -3.5) rectangle ++(3, 1) node[midway] {Verified};

      \draw[fill=red] (6, -3.5) rectangle ++(3, 1) node[midway] {Counterexample};
      \end{tikzpicture}
    \end{adjustbox}
  \end{minipage}
  \begin{minipage}{.38\linewidth}
    \caption{
      The user writes down a $\BehaviorTree$ along with specifications using the DSL\@.
      Depending on the desired output, a different mode of \textsf{BehaVerify} is used\@.
      It is possible to produce Python and Haskell implementations of the $\BehaviorTree$\@.
      These implementations allow for execution of the $\BehaviorTree$, for example, in robotics systems via PyTrees and robot operating system (ROS)\@.
      A \texttt{.smv} model can be generated, which can then be used with \textsf{nuXmv}~\cite{nuXmv} to verify the $\BehaviorTree$ satisfies specifications\@.
      Should it fail, a counterexample will be produced\@.
    }\label{fig:behaverify}
  \end{minipage}

\end{figure}


\subsection{Domain Specific Language (DSL) Input}
The input for \textsf{BehaVerify} is provided using a DSL: this includes the $\BehaviorTree$, the environment, and any specifications to be verified\@.
We used \textsf{textX}~\cite{textX} to create a grammar for $\BehaviorTree$s\@\footnote{The \textsf{textX} file that defines the grammar can be found at: ~\url{https://github.com/verivital/behaverify/tree/main/src/behaverify/data/metamodel}.}\@.
Some specific features of \textsf{BehaVerify} relating to the DSL are as follows:
\textbf{Functions:}
  Functions are written in a Lisp style\@.
  For example, $1+2$ would be \texttt{(add, 1, 2)}\@.
\textbf{Nondeterminism:}
  Blackboard variables can be updated nondeterministically between time steps, or each tick for environment variables.
  If more than one value is in a result, a value will be chosen nondeterministically\@.
\textbf{Loops:}
  Loops are unraveled by \textsf{BehaVerify}\@.
  The example loops are used to assign values nondeterministically\@.
\textbf{Types:}
  Variables are Booleans, enumerations or bounded integer ranges such as \texttt{[0, 19]}, which become finite-domain variables in \textsf{nuXmv}\@.
  The DSL also accepts unbounded \texttt{INT} and \texttt{REAL} variables, which become \textsf{nuXmv} \texttt{integer} and \texttt{real} variables\@.
  The verification modes of \toolname run \textsf{nuXmv}'s BDD-based engine, which needs finite domains; a model with unbounded variables must instead be checked with \textsf{nuXmv}'s SMT-based commands, which support invariants and LTL but not CTL\@.
  Every model in this paper uses bounded domains, so every reported result is for a finite-state system\@.
\textbf{Arrays:}
  \textsf{BehaVerify} supports array variables, though no arrays are featured in the example\@.
\textbf{Arguments:}
  Nodes can take arguments.
  To use an argument, modify the node to read \texttt{arguments\{arg {:}= INT\}}\@.
  In the tree, to pass \texttt{3} to a node labeled \texttt{Node}, use \texttt{Node\{3\}}.
\textbf{Specifications:}
  Specifications can be written using Linear Temporal Logic (LTL), Computation Tree Logic (CTL), and Invariant specifications\@.

To show how the DSL is actually used, Listing~\ref{lst:drone-dsl}
gives the complete encoding of the drone $\BehaviorTree$ of
Figure~\ref{fig:drunk-drone}, including its CTL specifications. The
specifications read `\BlVar{x\_g} at 0' because of fastforwarding: `at $n$'
selects the stage of the variable within the tick, as described in Section~\ref{subsec:ff}.

\begin{lstlisting}[language=BehaVerify, style=BehaVerifyStyle, upquote=True, basicstyle=\small\ttfamily, caption={The complete \toolname DSL encoding of the drone example in Figure~\ref{fig:drunk-drone}.}, captionpos=b, label={lst:drone-dsl}]
configuration {} enumerations {'We', 'Ea', 'No', 'So', 'XX'}
constants {min_val := 0, max_val := 19}
variables {
    variable { bl act VAR {'We', 'Ea', 'No', 'So', 'XX'} assign{result{'XX'}}}
    variable { bl new VAR BOOLEAN assign{result{False}}}
    variable { env x_d VAR [min_val, max_val] assign{result{min_val}}}
    variable { env y_d VAR [min_val, max_val] assign{result{min_val}}}
    variable { env x_g VAR [min_val, max_val] assign{result{x_d}}}
    variable { env y_g VAR [min_val, max_val] assign{result{y_d}}}
} environment_update {
    variable_statement { x_g #{if new, assign value nondeterministically, else leave unchanged}#
	assign { case {new} result {(loop, loop_var, [min_val, max_val] such_that True, loop_var)}
	    result {x_g}}}
    variable_statement {y_g
	assign {case{new} result {(loop, loop_var, [min_val, max_val] such_that True, loop_var)}
	    result {y_g}}}
    variable_statement {x_d
	assign {
	    case {(eq, act, 'We')} result { (max, min_val, (sub, x_d, 1))}
	    case {(eq, act, 'Ea')} result { (min, max_val, (add, x_d, 1))}
	    result {x_d}}}
    variable_statement {y_d
	assign {
	    case {(eq, act, 'So')} result { (max, min_val, (sub, y_d, 1))}
	    case {(eq, act, 'No')} result { (min, max_val, (add, y_d, 1))}
	    result {y_d}}}	    
} checks {}
environment_checks {
    environment_check { NeedNew
	arguments {} read_variables { x_d, y_d, act}
	condition {(and, (eq, x_g, x_d), (eq, y_g, y_d))}}
} actions {
    action { NewGoal
	arguments{} local_variables {}read_variables {} write_variables {new, act}
	initial_values {}
	update {
	    variable_statement { new assign{result{True}}}
	    variable_statement { act assign{result{'XX'}}}
	    return_statement {result {running}}
	}
    } action {NextAct
	arguments{} local_variables {}read_variables {} write_variables {new, act}
	initial_values {}
	update {
	    variable_statement {new assign{result{False}}}
	    variable_statement {act assign {result{'No', 'So', 'We', 'Ea'}}}
	    return_statement {result { success}}
	}
    }
} sub_trees {}
tree {
    composite {Drone selector children {
	    composite { GoalSeq sequence children { NeedNew {} NewGoal {}}}
	    NextAct {}
	}
    }
}
tick_prerequisite {(True)}
specifications {
    CTLSPEC {(always_globally, (always_finally, (and, (eq, x_d at -1, x_g at 0), (eq, y_d at 1, y_g at 0))))}
    CTLSPEC {(always_globally, (exists_finally, (and, (eq, x_d at -1, x_g at 0), (eq, y_d at 1, y_g at 0))))}
}
\end{lstlisting}

\subsection{Fastforwarding}
\label{subsec:ff}
A naive encoding of $\BehaviorTree$s into model checking languages would model each node activation as a separate transition\@.
This approach is problematic: a tree with 100 nodes would require 100+ transitions per tick, leading to extremely long traces and poor verification performance\@.
As demonstrated in~\cite{serbinowska2024FMAS-SBT}, BehaVerify uses \textbf{Fastforwarding} to compress an entire tick into a single transition\@.
Rather than modeling the step-by-step traversal, we compute the final state after a complete tick in one transition\@.
This dramatically reduces trace lengths and is a primary reason \textsf{BehaVerify} outperforms other tools in verification time\@.
Table~\ref{fig:drunk-drone-table} illustrates the difference. \toolname's naive encoding (\texttt{--use\_encoding naive}) takes one step per node visit plus one step between ticks, which is one state per column, so the two ticks shown take 15 transitions. Fastforwarding keeps only the initial state and the state at the end of each tick, $s_0$, $s_1$ and $s_2$ in the row \textbf{FF}, so each tick is a single transition and a counterexample trace has one step per tick.

\paragraph{Relation to Existing Encodings.}
Fastforwarding applies a well-known idea rather than a new one. A tick is a loop-free computation: each node runs at most once, in an order fixed by the tree. Summarizing such a computation as one transition, with a fresh version of each variable after every assignment and intermediate values given by definitions instead of state variables, is the static single assignment (SSA) form~\cite{cytron1991TOPLAS}. SSA-style symbolic encodings are standard in software verification, for example large-block encoding in software model checking~\cite{beyer2009FMCAD} and bounded model checking of C programs~\cite{clarke2004TACAS}. The stages described below are the SSA versions of the blackboard and environment variables. What is specific to $\BehaviorTree$s is how the guards of these versions arise and how they are exposed: return statuses decide which nodes run in a tick, nondeterministic choices inside a tick remain branching of the single transition, and stages stay visible to specifications through \texttt{at}, so a user can still state properties of values inside a tick.

\paragraph{Stages.}
A variable can change value during a tick, so we need a way to refer
to it at different points within a single tick; we call these \emph{stages}\@.
\emph{Stage $0$} is the value at the start of the tick, \emph{Stage $1$} the
value after the first possible change, and so on, with \emph{Stage $-1$}
denoting the last stage\@. Here `the first possible change' means the first node
able to change that variable under a depth-first traversal; if that node is
skipped, the stage falls back to the value of the previous stage\@.

A natural concern is \emph{state space explosion}: if a variable has 10 possible
values and we keep three stages, that appears to be 1000 possible values\@. There
is a simple way around it. If a variable updates deterministically then each
stage is a function of the previous one, so the stage can be stored as a
\texttt{DEFINE} and costs no state at all\@. If, for instance, $x$ is updated by
incrementing it, we need no new variable for the new stage --- we simply insert
$x+1$ wherever it is needed\@. In some cases a stage can be removed outright\@.
Specifications do not need to be rewritten when using fastforwarding: the LTL
next operator ($X$) still refers to the next tick, \textbf{not} the next node
activation~\cite{serbinowska2024FMAS-SBT}\@.

\paragraph{Why Fastforwarding Is Sound.}
A compression of this kind can be mistaken for an abstraction that would require
a preservation proof, so it is worth being precise. Fastforwarding is \emph{not}
an abstraction of a finer model. It directly constructs the tick-level
transition relation that \emph{defines} $\StatefulBehaviorTree$ semantics
in~\cite{serbinowska2024FMAS-SBT}, whose states are exactly the blackboard and
environment valuations at tick boundaries. Deterministic intra-tick updates fold
into \texttt{DEFINE}s, as above, and each nondeterministic intra-tick choice is
exposed as a choice of that same transition. The resulting Kripke structure is
identical to the tick-level semantics up to auxiliary variables that no
specification refers to --- the two are bisimilar --- and bisimulation preserves
full CTL$^*$. Branching is preserved because each tick's nondeterminism becomes
branching at a single transition, and intra-tick values remain expressible
through stages. Empirically, both encodings returned identical verdicts on every
instance where the naive encoding completed
(Table~\ref{tab:ff-ablation}).

\paragraph{State-Space Growth.}
In the fastforwarded model, the activity and return status of every node is a \texttt{DEFINE} computed from the stages. A state therefore holds only the blackboard and environment values at the start of a tick, plus one variable for each nondeterministic choice inside a tick; there is no program counter and no per-node status. The size of the state space depends on the variables and their domains, not on the number of nodes. In the binary-tree benchmark of Table~\ref{tab:ff-ablation}, the blackboard is a counter $vv \in [0, L]$, where $L = 2^N$ is the number of leaves in the scaled subtree, and a Boolean $rv$ with three stages, so the model has $8(L+1)$ states (at $N{=}10$, $\log_2(8 \cdot 1025) = 13.0$). Here the state count grows linearly with the number of nodes, $7 + 2^{N+1}$, only because this benchmark ties the range of $vv$ to the number of leaves. The naive encoding instead adds a program counter over all nodes and a three-valued status variable for every leaf. Its total state space therefore grows exponentially with the number of nodes, and its reachable states grow roughly quadratically in this benchmark.

\subsection{From Stateful to Neuro-Symbolic Behavior Trees}
\label{subsec:sbt-to-nsbt}
An $\NeuroSymbolicBT$ is an $\StatefulBehaviorTree$ in which some leaf nodes
consult a neural network. The step from one to the other is deliberately small:
the DSL gains declarations for neural networks and for the variables they read
and write, while everything else --- tick
semantics, blackboard, environment, specification language, fastforwarding ---
is unchanged. This is what lets $\NeuroSymbolicBT$s be verified by exactly the
same pipeline, with no separate front end.
A network declaration names an ONNX file, the input variables and the output classes, and selects how the network enters the \textsf{nuXmv} model. In \texttt{table} mode, \toolname runs the network with ONNX Runtime on every combination of input values, which is possible because network inputs must have bounded integer domains, and writes the predicted class as a \texttt{case} expression. The model stays finite-state and exact for the given network. \toolname also offers symbolic modes that write each \texttt{Gemm} and \texttt{ReLU} layer as piecewise-linear definitions~\cite{albarghouthi2021introduction}, over reals or over fixed-width bit vectors; the experiments in this paper use \texttt{table} mode.

\subsection{Output Targets}
\textsf{BehaVerify} can output Python code, Haskell code, or a \texttt{.smv} model for verification with \textsf{nuXmv}\@.
In the case of Python and Haskell code, \textsf{BehaVerify} creates the necessary tree structure and appropriately sets up variables\@.
A simulation interface is also formed; this will run the code and print out relevant information at the end of each tick alongside an environment that evolves according to the rules in the input file provided by the user\@.
In the Python implementation, the environment file can be treated as an interface through which the user connects the $\BehaviorTree$ to the true environment it is being used in\@.
For instance, a more practical version of the drone in Figure~\ref{fig:drunk-drone} might include an extra leaf node `SendAct' that uses a write environment statement in its update to communicate with the environment\@.
The generated Python leaf node would include a function call, to a function in the environment, that the user could overwrite to implement sending an action in practice (\emph{e.g.}, through  \textsf{ROS2}~\cite{ROS2})\@.
As we use \textsf{PyTrees} for our Python implementation, we have deployed numerous $\BehaviorTree$s with specifications verified from our DSL in \textsf{nuXmv} to corresponding implementations in several robotic systems, including underwater vehicles and quadcopter drones~\cite{serbinowska2022SEFM,serbinowska2025NEUS-REPLACE}, via the ROS PyTree library\footnote{\url{https://index.ros.org/p/py_trees/}}\@.
\par{}
The two code backends serve different purposes. Python targets
\emph{deployment}: it builds on \textsf{PyTrees} and ROS, and has been run on
real underwater and quadcopter systems. Haskell provides an \emph{executable
reference semantics} --- pure, typed code mirroring the formal tick function
directly --- which we use for differential testing against both the Python
backend and the \textsf{nuXmv} model, at compiled speed. The generated Haskell is
deliberately self-contained, depending only on \texttt{base}, \texttt{random},
\texttt{array} and \texttt{containers}, which avoids inheriting the semantic gaps
of third-party $\BehaviorTree$ libraries (compare the \textsf{PyTrees} versus
\texttt{BehaviorTree.cpp} discussion in Section~\ref{subsec:BT2Fiacre}, or
Haskell's \textsf{smarties}). C, C++ and Rust backends are planned.

\textsf{nuXmv} can be used to verify \emph{Linear Temporal Logic} (LTL), \emph{Computation Tree Logic} (CTL), and Invariant specifications provided by the user\@.
In our example, a CTL specification can state that the drone will \emph{always eventually} reach the goal; this claim is false and \textsf{nuXmv} provides a counterexample\@.
To understand why, consider that the drone moves
nondeterministically, selecting North, South, East or West each tick\@.
A counterexample trace might show the drone oscillating North-South indefinitely, never reaching the goal\@.
Another specification can state that the drone is \emph{always capable of eventually} reaching the goal; \textsf{nuXmv} confirms this claim is true\@.
The key difference is ``always eventually'' ($AF$) versus ``always
exists a path to eventually'' ($AGEF$)\@. $AF$ requires \emph{all} paths to
reach the goal, which fails under nondeterministic movement\@. $AGEF$ requires
only that from any state \emph{some} path exists that eventually reaches the
goal, and since the drone can always choose the correct direction, such a
path always exists\@.
\subsection{Visualization and Graphical User Interface}
\label{subsec:GUI}

\begin{figure}[t]
\includegraphics[width=\linewidth,keepaspectratio]{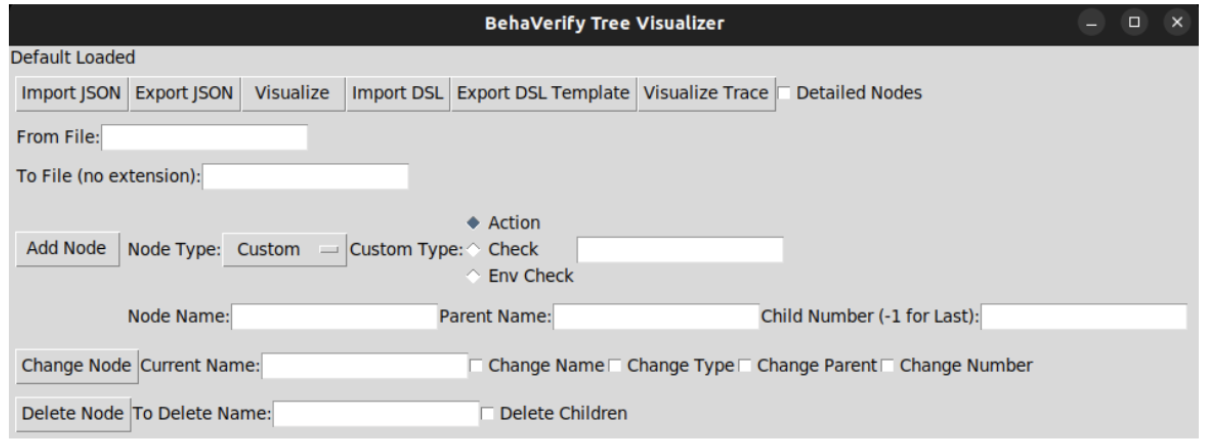}
    \caption{Graphical User Interface for \textsf{BehaVerify}}
    \label{fig:gui}
\end{figure}

\begin{figure}[t]
\centering
\includegraphics[width=0.8\linewidth,keepaspectratio]{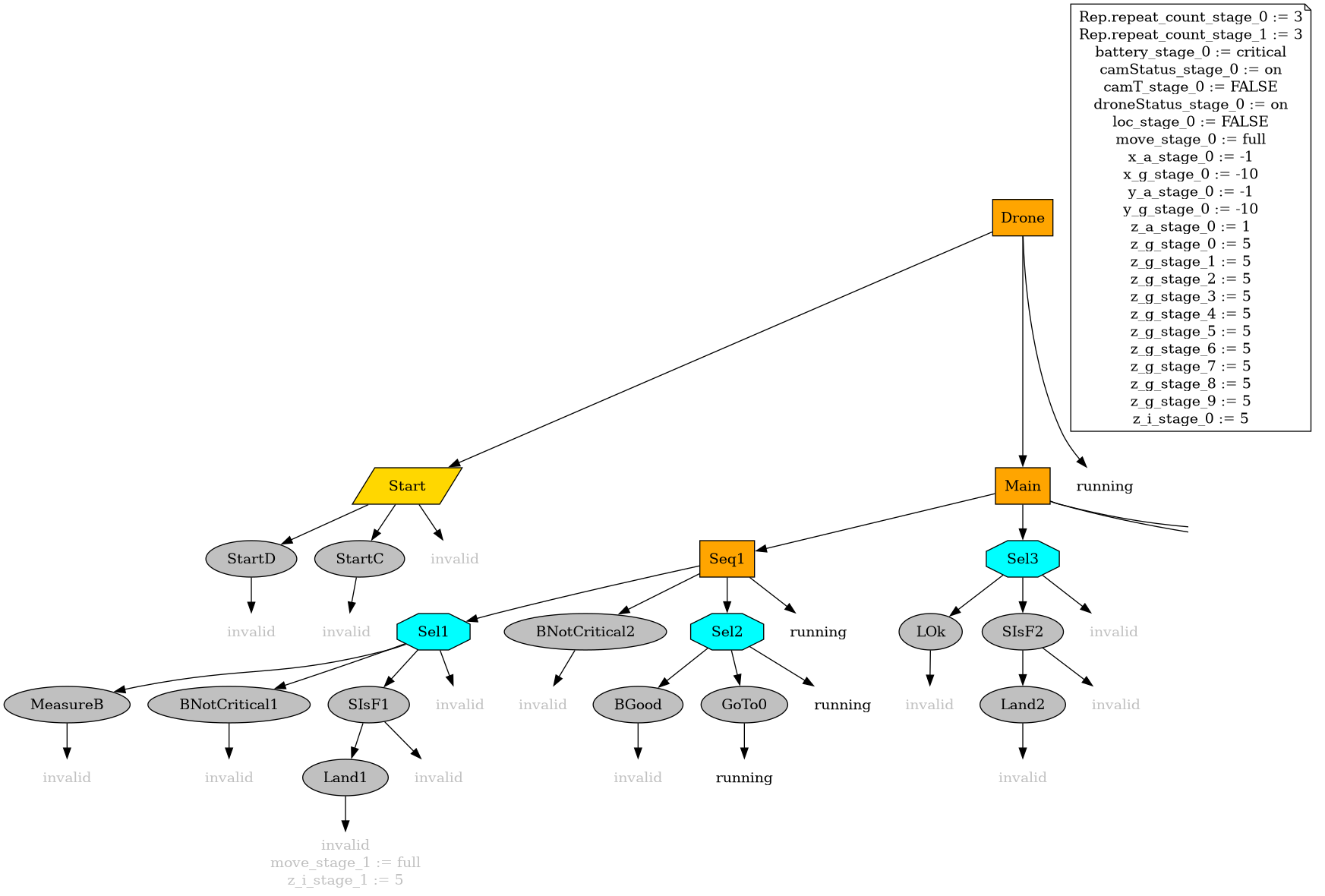}
\caption{One tick of the \texttt{drone3} counterexample of Section~\ref{subsec:BT2Fiacre}, drawn by \texttt{behaverify trace} (cropped). The battery is \texttt{critical}, but \texttt{Seq1} resumes its running child \texttt{Sel2}, so the battery checks under \texttt{Sel1} are skipped and \texttt{Land1} stays \texttt{invalid}.}
\label{fig:trace}
\end{figure}

\toolname includes a graphical user interface (GUI), launched with \texttt{behaverify gui} and shown in Figure~\ref{fig:gui}. It is written in Python with Tkinter and draws trees with Graphviz. A user can import a tree from a \texttt{.tree} file or from JSON, add, change and delete nodes, and render the tree, optionally labeling every leaf with its type. The edited tree can be exported as JSON or as a \texttt{.tree} skeleton that holds the tree structure and stub leaves; the user then adds variables, leaf behavior and specifications in the DSL. It also displays counterexample traces, but does not yet edit leaf behavior or run verification, which remain DSL and command-line steps.

\paragraph{Tree Visualization.}
\toolname's \LaTeX{} renderer is separate from the GUI: \texttt{behaverify latex} turns a \texttt{.tree} file into a TikZ diagram that also shows the code of every leaf, the initial values, the variable domains and the environment updates. Figure~\ref{fig:drunk-drone} and the other tree diagrams in this paper were drawn this way.

\paragraph{Trace Visualization.}
\textsf{nuXmv} produces counterexamples when a specification is falsified\@.
This result is valuable standalone, but also enables a user to correct any mistakes present in the model\@.
This requires a familiarity with both \textsf{nuXmv} and the \textsf{BehaVerify} encoding in \textsf{nuXmv} --- for particularly large examples the process of reading a trace may be tedious\@.
To alleviate this, we developed the ability to visualize \textsf{nuXmv} traces: \texttt{behaverify trace}, also available in the GUI, turns a counterexample into one Graphviz image per tick (Figure~\ref{fig:trace}). Each node is labeled with the status it returned in that tick, colored by status; each leaf lists the staged values it wrote; and a side box lists the values at the start of the tick. Because fastforwarding makes each tick a single step, a counterexample of $k$ ticks becomes $k$ images, whatever the size of the tree. For grid-world models, \texttt{behaverify grid} also draws the trajectory of the agent as an image and an animation.
\subsection{Runtime Monitoring}
\textsf{BehaVerify} also has support for \emph{Behavior Trees with Monitor} ($\BehaviorTreeMonitor$)~\cite{serbinowska2024FMAS-BTM}. $\BehaviorTreeMonitor$ is a variant $\BehaviorTree$ model that allows for the creation of monitors using LTL specifications that were embedded in leaf nodes. This allows for detection of liveness violations, and can be used to monitor whether a $\BehaviorTree$ makes progress at runtime. In the interest of space, we do not describe this further here, but interested readers can see details in~\cite{serbinowska2024FMAS-BTM}\footnote{\url{https://github.com/verivital/behaverify/tree/main/REPRODUCIBILITY}}\@.

\subsection{\texorpdfstring{Testing and Distribution}{Testing and Distribution}}
\label{subsec:ci}
\toolname is released on PyPI (version 1.0.0) by a GitHub Actions workflow that builds and publishes each tagged release. A second workflow runs on every push and pull request. It runs the 406 \texttt{pytest} tests on Python 3.10, 3.11 and 3.12; these cover the command-line interface and every mode, parsing example models, and generation and execution of Python code for example trees. It also runs each mode of the command-line tool on example models, and it downloads \textsf{nuXmv} and checks that verification completes on three example models. A third workflow builds the documentation and publishes it to GitHub Pages.

\section{Comparisons and Results}
\label{sec:comparisons}
We compare \toolname against \textsf{MoVe4BT}~\cite{MoVe4BT}, \textsf{BT2Fiacre}~\cite{ingrand2025arXiv}, and \textsf{BT2BIP}~\cite{wang2024FACS}, and contrast \toolname's fastforwarding encoding to a naive \textsf{nuXmv} encoding. Trees, \texttt{.smv} models, raw timings (including reachable/total state-space counts across all $N$), and counterexample traces are in our reproducibility package. All experiments used a 6-core, 12-thread i7-10750H Intel CPU with a 16GB RAM. 

\paragraph{On the design of this evaluation.}
An all-tools-on-all-benchmarks matrix is not achievable ; no two $\BehaviorTree$ tools share an input format or node
semantics: a \textsf{PyTrees} sequence is a \texttt{BehaviorTree.cpp}
\emph{reactive} sequence, and \texttt{BehaviorTree.cpp}'s sequence-with-memory
has no \textsf{PyTrees} counterpart at all. Building the other toolchains also
took substantial effort, and some we could not run despite dockerized,
documented attempts and issues opened upstream. We therefore compare on each
tool's own published examples --- the setting most favorable to that tool.
\toolname installs through \texttt{pip} or Docker and reproduces
its results end to end, which was not our experience elsewhere.

\paragraph{Reproducibility.}
The comparison setups are public. Our reproducibility package~\cite{behaverify2026zenodo} contains one Docker image that builds \toolname, \textsf{nuXmv} and the \textsf{BT2Fiacre} toolchain (\textsf{Fiacre}, \textsf{Hippo} and \textsf{TINA}) at pinned versions, and regenerates Tables~\ref{tab:ff-ablation} and~\ref{tab:drone3compare} and the \textsf{BT2BIP} verdicts. The recreated trees are in the \toolname repository. \textsf{MoVe4BT} has only a graphical interface, so we provide its generated inputs, a containerized desktop in which it runs, and its recorded logs. The released \textsf{BT2BIP} toolchain did not run in our containers, so only the \toolname side of that comparison is reproduced.

\begin{table}[t]
\centering
\small
\setlength{\tabcolsep}{4pt}
\begin{tabular*}{\linewidth}{@{\extracolsep{\fill}}l c c c c}
\toprule
\textbf{Capability} & \textbf{\toolname} & \textbf{MoVe4BT} & \textbf{BT2Fiacre}
  & \textbf{BT2BIP} \\
\midrule
LTL specifications          & \checkmark & \checkmark & \checkmark        & \checkmark \\
CTL specifications          & \checkmark & --         & $\checkmark\ddagger$ & --  \\
Invariant checking          & \checkmark & --         & \checkmark        & \checkmark  \\
Variables in specifications & \checkmark & --         & \checkmark        & \checkmark  \\
Non-Boolean variables       & \checkmark & \checkmark & \checkmark        & \checkmark \\
Neural-network leaves       & \checkmark & --         & --                & --  \\
Runtime monitors            & \checkmark & --         & \checkmark$\S$ & --  \\
True parallel nodes         & --         & \checkmark & \checkmark        & \checkmark \\
Time-bounded specifications & $\dagger$  & \checkmark & $\checkmark$ & --  \\
Code generation             & \checkmark & --         & \checkmark$\S$ & --  \\
\texttt{pip}/Docker install & \checkmark & --         & --$\P$       & --  \\
\bottomrule
\end{tabular*}
\caption{Feature comparison across $\BehaviorTree$ verification tools.
$\dagger$: \toolname has no per-node duration primitive, but time-bounded
properties are recoverable through a tick-counter blackboard variable.
$\ddagger$: supported by the underlying Fiacre/TINA formalism.
$\S$: through the \textsf{Hippo} engine, which compiles a Fiacre model to executable C and supports run-time verification~\cite{hladik2021JSS,dalzilio2023RAS}.
$\P$: the \textsf{BT2Fiacre} toolchain is distributed as source and binaries; its \textsf{TINA} model checker is binary-only.
``\checkmark'' and ``--'' denote supported and not supported, respectively.}
\label{tab:feature-comparison}
\end{table}

\paragraph{Limitations.}
Our comparisons inherit the semantic mismatch described above, so a recreated
tree is the closest faithful equivalent rather than an identical model.
\toolname follows \textsf{PyTrees} semantics and so does not offer true parallel
node execution, which \textsf{BT2BIP} and \textsf{MoVe4BT} do; this is a
deliberate alignment with the library we deploy against rather than an
implementation gap. \toolname also has no native notion of per-node duration.
Finally, the \textsf{BT2BIP} examples are small enough that timing comparisons
against them carry little information.

\paragraph{Neuro-Symbolic Behavior Trees.}
\toolname supports $\NeuroSymbolicBT$s, trees with neural-network leaf nodes, following the theory of~\cite{serbinowska2025NEUS-REPLACE}.
A neural
network verifier checks open-loop input--output properties of an isolated
network~\cite{albarghouthi2021introduction}. $\NeuroSymbolicBT$ verification instead checks temporal properties of
the closed loop formed by the tree, the network and a nondeterministic
environment. To our knowledge no other $\BehaviorTree$ tool supports
neural-network leaves at all.

We report results for a drone that seeks to avoid obstacles on a $10\times 10$ grid with $22$ obstacles while navigating to randomly assigned goals\@. This is the $\BehaviorTree$
of Figure~\ref{fig:drunk-drone}, modified so that a neural network replaces the
A* search algorithm\@. We train two fully-connected networks, each with one layers of 1000 neurons and ReLU activations\@. The input is the current
position and the goal position (four integers) and the output is one of five actions: the four
directions or no move\@. One network reached 1.0 accuracy on the training data, the other
0.9995\@. Both networks are used in \texttt{table} mode. \toolname evaluates each network on all $10^4$ combinations of its four inputs and stores the chosen action as a 10\,000-row \texttt{case} expression, so the model has no real-valued variables and the hidden layer never reaches the model checker. \textsf{nuXmv} then checks the specifications with its standard BDD-based algorithms (\texttt{check\_invar} and \texttt{check\_ctlspec}). No abstraction is involved, so the verdicts are exact for these networks. The cost grows with the size of the input domain, which determines the number of table rows, rather than with the size of the network; larger or continuous input domains need the symbolic modes instead. For each network we attempt to verify an invariant and a CTL specification:
\begin{equation*}
  \begin{split}
    \text{INVAR: } \quad&(\EnvVar{x\_d}, \EnvVar{y\_d}) \not\in Obs\\
    \text{CTL: } \quad&AG(AF((\EnvVar{x\_d}, \EnvVar{y\_d}) = (\EnvVar{x\_g}, \EnvVar{y\_g})) \lor (\EnvVar{x\_g}, \EnvVar{y\_g}) \in Obs
  \end{split}
\end{equation*}
where $Obs$ is the set of obstacles\@. In both cases, the open-loop, obstacle-avoidance invariant was verified\@.

The closed-loop CTL specification, however, could only be verified for the perfect-accuracy network. For the 0.9995-accuracy network, the counter-example trace repeatedly emitted a non-progressing action over a multi-tick trajectory, so the drone stopped and never reached the target. This closed-loop behavior cannot simply be expressed as an open-loop query about the network alone\@.

\restored{\begin{table}[t]
\centering
\small
\begin{tabular*}{\linewidth}{@{\extracolsep{\fill}}l c c}
\toprule
\textbf{Task} & \textbf{1.0000 Acc} & \textbf{0.9995 Acc}\\
\midrule
Translation Time & 0.55 & 0.54\\
Invariant Time   & 0.87 & 0.95\\
CTL Time         & 9.24 & 33.15\\
\bottomrule
\end{tabular*}
\caption{Timing results for the neural network example implemented as an
$\NeuroSymbolicBT$. All times in seconds, truncated to two decimal places
because that is the precision \textsf{nuXmv} reports. \textbf{Translation Time}
is the time to translate a \texttt{.tree} file to a \texttt{.smv} file.}
\label{tab:nsbt-timing}
\end{table}}

\begin{table}[t]
\centering
\small
\setlength{\tabcolsep}{4pt}
\begin{tabular*}{\linewidth}{@{\extracolsep{\fill}} r r  c c c c  c c }
\toprule
& & \multicolumn{2}{c}{\textbf{CTL (s)}} & \multicolumn{2}{c}{\textbf{LTL (s)}}
  & \multicolumn{2}{c}{\textbf{Reachable / Total ($\log_2$)}} \\
\cmidrule(lr){3-4}\cmidrule(lr){5-6}\cmidrule(lr){7-8}
$N$ & nodes & \textbf{FF} & \textbf{Naive} & \textbf{FF} & \textbf{Naive}
            & \textbf{FF} & \textbf{Naive} \\
\midrule
1  & 11   & 0.01 & 0.01    & 0.01   & 0.03    & 3.17 / 4.58  & 6.94 / 17.38 \\
2  & 15   & 0.01 & 0.02    & 0.02   & 0.08    & 4.39 / 5.32  & 8.16 / 21.67 \\
3  & 23   & 0.01 & 0.06    & 0.06   & 0.18    & 5.49 / 6.17  & 9.49 / 29.42 \\
4  & 39   & 0.01 & 0.14    & 0.13   & 0.56    & 6.54 / 7.09  & 10.95 / 43.73 \\
5  & 71  & 0.02 & 0.62    & 0.38   & 3.54    & 7.56 / 8.04  & 12.57 / 70.88 \\
6  & 135  & 0.06 & 4.62    & 1.31   & 35.49   & 8.57 / 9.02  & 14.32 / 123.48 \\
7  & 263  & 0.11 & 53.03   & 4.27   & TIMEOUT & 9.58 / 10.01 & 16.17 / 226.86 \\
8  & 519 & 0.24 & TIMEOUT & 13.62  & TIMEOUT & 10.58 / 11.01 & TIMEOUT \\
9  & 1031 & 0.57 & TIMEOUT & 45.75  & TIMEOUT & 11.58 / 12.00 & TIMEOUT \\
10 & 2055 & 2.24 & TIMEOUT & 153.65 & TIMEOUT & 12.58 / 13.00 & TIMEOUT \\
\bottomrule
\end{tabular*}
\caption{Fastforwarding-vs-naive ablation on a binary-tree benchmark scaled by depth $N$ (5\,min per-instance timeout, times in seconds, state-space columns as $\log_2$). The timing results presented here are the averages calculated over three separate runs. \textbf{FF} is \toolname's fastforwarding encoding. \textbf{Naive} is the same tool with \texttt{--use\_encoding naive}. Under \textbf{FF}, reachable and total states grow linearly with the number of nodes; under Naive, reachable states grow roughly quadratically and the total state space exponentially. For example, at $N{=}7$ the naive total state space is already $2^{226}$ versus FF's $2^{10}$. Naive timed out at $N{\geq}8$ for both verification and reachability computation.}
\label{tab:ff-ablation}
\end{table}

\subsection{Fastforwarding vs.\ Naive Encoding}
The scalability advantage over \textsf{MoVe4BT} (50--20\,000 nodes, approx. two orders of magnitude) and the simple-robot blackboard-scaling result (grids up to $30{\times}30$) were established in~\cite{serbinowska2024FMAS-SBT}. \textsf{MoVe4BT} additionally lacks invariant checking, CTL and variables-in-specifications; its only edge is per-node duration for time-bounded LTL, recoverable in \toolname through a tick-counter blackboard variable. Here we add what was \emph{not} ablated in~\cite{serbinowska2024FMAS-SBT}: fastforwarding versus a naive \textsf{nuXmv} encoding within \toolname itself. The ablation in Table~\ref{tab:ff-ablation}, on a binary-tree benchmark scaling to 2\,055 nodes, confirms that fastforwarding is what enables linear growth in reachable states, against roughly quadratic growth in reachable states for the naive encoding.

\restored{\begin{figure}[t]
  \centering
  \includegraphics[width=0.95\linewidth,keepaspectratio]{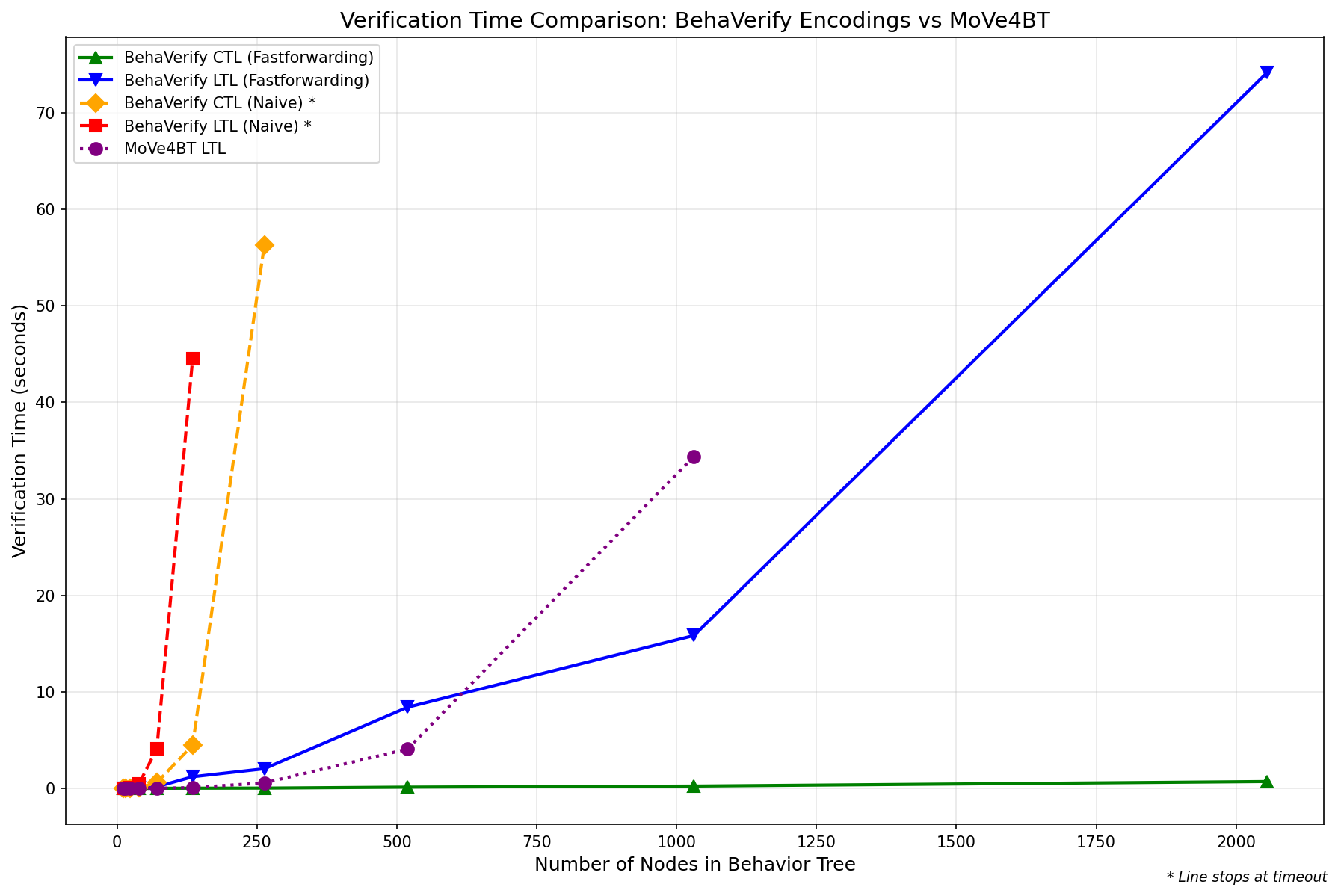}
  \caption{Timing for the binary-tree experiment, comparing \toolname's naive
  encoding, \toolname's fastforwarding encoding, and \textsf{MoVe4BT}, with a
  5-minute timeout\@. The node count grows as $7+2^{N+1}$, from 11 to 2\,055\@.
  At 2\,055 nodes \textsf{MoVe4BT} reported
  \texttt{QProcess: Destroyed while process ("mono") is still running}\@. Times
  exclude loading time for \textsf{MoVe4BT} and translation time for
  \toolname; the latter is under one second except at the largest size, where
  it is three seconds\@.}
  \label{fig:move4bt-timing}
\end{figure}}

Figure~\ref{fig:move4bt-timing} plots the same benchmark against
\textsf{MoVe4BT} as well, placing the encoding ablation of
Table~\ref{tab:ff-ablation} in context: the gap between \toolname's two
encodings is of the same order as the gap between \toolname and an entirely different
tool\@.

\subsection{BT2Fiacre: Drone}\label{subsec:BT2Fiacre}
\toolname follows \textsf{PyTrees} semantics whereas  \texttt{BehaviorTree.cpp}~\cite{BehaviorTreeCPP} is used by \textsf{BT2Fiacre} (e.g., a PyTrees sequence is a \texttt{BehaviorTree.cpp} reactive sequence; \texttt{BehaviorTree.cpp}'s sequence-with-memory has no direct PyTrees counterpart), so an exact recreation is not possible. We recreate the UAV example of~\cite{ingrand2025arXiv} as \texttt{drone3} (closest possible) and \texttt{droneNew} (\toolname rewrite that moves state from memory-bearing nodes onto the blackboard). Both trees, the integer-grid encoding ($x,y\in[-30,30]$, height in $[0,80]$), and the height-bound invariant are in the repository.\footnotemark[\value{footnote}] Table~\ref{tab:drone3compare} shows \toolname is several orders of magnitude faster than \textsf{BT2Fiacre}, which times out when trying to verify the height invariant. \texttt{droneNew} further shrinks the reachable state space from $2^{22}/2^{51}$ to $2^{16}/2^{40}$.

\begin{table}[t]
\centering
\small
\begin{tabular*}{\linewidth}{@{\extracolsep{\fill}}l c c c c}
\toprule
& \multicolumn{2}{c}{\textbf{\toolname}} & \multicolumn{2}{c}{\textbf{BT2Fiacre}} \\
\cmidrule(lr){2-3}\cmidrule(lr){4-5}
\textbf{Task} & \textbf{drone3} & \textbf{droneNew} & \textbf{drone3} & \textbf{drone4 (droneNew)} \\
\midrule
Prep.\                    & 0.73  & 1.14  & 0.21    & 0.24    \\
Check Height              & 86.87 & 0.48  & TIMEOUT & TIMEOUT \\
Reach.\ states ($\log_2$) & 22.83 & 16.53 & TIMEOUT & TIMEOUT \\
Total states ($\log_2$)   & 51.52 & 39.97 & TIMEOUT & TIMEOUT \\
\bottomrule
\end{tabular*}
\caption{Drone (\textsf{BT2Fiacre} comparison). Times in seconds, averaged over three separate runs. Timeout set to 5 minutes. ``Prep.''\ is \texttt{.tree}$\to$\texttt{.smv} translation for \toolname; tree-loading for \textsf{BT2Fiacre}. ``Check Height'' verifies the height stays within bound. BT2Fiacre's \texttt{drone4} is a reactive reformulation analogous to BehaVerify's \texttt{droneNew}. \texttt{droneNew} restructures \texttt{drone3} into a reactive style, shrinking reachable states and total states by approximately $2^{6}$ and $2^{12}$ respectively.}
\label{tab:drone3compare}
\end{table}

\toolname additionally falsifies a battery-priority invariant ($Main_{act} \implies ((battery{=}\textsf{`critical'}) \implies Land_{1,act})$) on \texttt{drone3}: because the tree relies on memory-bearing nodes, a \emph{low}-battery navigation event can mask a subsequent \emph{critical} event, and the drone fails to land. This is not an expressiveness gap on \textsf{BT2Fiacre}'s part: the flaw is
in the tree and is expressible in their framework, so Fiacre should in principle
falsify it too, resources permitting (it timed out on the simpler height
invariant). The point is rather that cheap specifications plus fast verification
meant more properties actually got checked, which is how the bug surfaced at
all. \textsf{BT2Fiacre} did not flag it because their original LTL
specification did not require checking the battery. Figure~\ref{fig:trace} shows the tick in which the property fails; the full counterexample trace is in the repository\@.

\subsection{BT2BIP}
We recreated the Mars Rover and Train Control examples of~\cite{wang2024FACS} (originally from~\cite{biggar2020IEEE-RAL}) in \toolname. Both tools agree the Mars Rover spec $\EnvVar{storm} \implies (Hibernate_{st}=\Success)$ is false (low power preempts hibernation) and the Train Control spec $(\BlVar{s}\implies\BlVar{v}{=}0)\land(\BlVar{d}\implies\BlVar{v}{=}0)$ is false ($\BlVar{s}$ is never reset). \textsf{BT2BIP} timed out on LTL specifications on the larger trees while \toolname completed verification.

\section{Conclusions and Future Work}
\label{sec:future-work}
We presented \textsf{BehaVerify}, a tool for formal verification of Behavior Trees\@. We highlighted the tool's maturity through the contributions of Neuro-Symbolic Behavior Trees and Stateful Behavior Tree support, and the ability to formalize behavior trees in the Domain Specific Language (DSL) through verification and implementation. \textsf{BehaVerify} is scalable as shown via a 2\,055 node case study comparison, illustrating \toolname's superior scalability to existing methods due to the efficient fastforwarding encoding into \textsf{nuXmv}, versus naive encodings and those used in some other behavior tree verification tools.
Future work includes plans to expand the output options for \textsf{BehaVerify} with other frameworks, such as \texttt{BehaviorTree.cpp}\@.
The DSL is verbose for end users, so we plan to let the GUI edit leaf behavior and specifications and launch verification, to offer specification patterns or a higher-level syntax for temporal properties, and to run \toolname as a language server so that an editor such as VS Code can check a tree while it is being designed. We also plan to compare \toolname with \textsf{BT2Fiacre} on a simulated robotic application.
We also plan to continue developing and improving the tool overall, as we are using it in several ongoing projects for autonomous systems and robotics applications. We will also explore physical instantiations of the self-correcting neuro-symbolic framework derived in~\cite{wooding2026selfcorrecting}.

\section*{AI Usage}

Generative AI was used to assist with polishing the manuscript; including checking that results presented match the underlying code and validating the accuracy of the features in Table 2 (each also manually checked).

\section{Acknowledgments}
The material presented in this paper is based upon work supported by the National Science Foundation (NSF) through grant number 2220401, the Defense Advanced Research Projects Agency (DARPA) under contract number FA8750-23-C-0518, and the U.S. Department of Energy, Office of Science, Office of Advanced Scientific Computing Research, under Award Number DE-SC0025528. Any opinions, findings, and conclusions or recommendations expressed in this paper are those of the authors and do not necessarily reflect the views of DARPA, DOE, or NSF.

\bibliographystyle{eptcs}
\bibliography{ATVARefs}

\begin{thebibliography}{10}
\providecommand{\bibitemdeclare}[2]{}
\providecommand{\surnamestart}{}
\providecommand{\surnameend}{}
\providecommand{\urlprefix}{Available at }
\providecommand{\url}[1]{\texttt{#1}}
\providecommand{\href}[2]{\texttt{#2}}
\providecommand{\urlalt}[2]{\href{#1}{#2}}
\providecommand{\doi}[1]{doi:\urlalt{https://doi.org/#1}{#1}}
\providecommand{\eprint}[1]{arXiv:\urlalt{https://arxiv.org/abs/#1}{#1}}
\providecommand{\bibinfo}[2]{#2}

\bibitemdeclare{article}{albarghouthi2021introduction}
\bibitem{albarghouthi2021introduction}
\bibinfo{author}{Aws \surnamestart Albarghouthi\surnameend}
  (\bibinfo{year}{2021}): \emph{\bibinfo{title}{Introduction to neural network
  verification}}.
\newblock {\slshape \bibinfo{journal}{Foundations and Trends{\textregistered}
  in Programming Languages}} \bibinfo{volume}{7}(\bibinfo{number}{1--2}), pp.
  \bibinfo{pages}{1--157}, \doi{10.1561/2500000051}.

\bibitemdeclare{misc}{BehaviorTreeCPP}
\bibitem{BehaviorTreeCPP}
\bibinfo{author}{\surnamestart {Auryn Robotics}\surnameend}:
  \emph{\bibinfo{title}{Tutorial 02: Blackboard and Ports}}.
\newblock
  \urlprefix\url{https://www.behaviortree.dev/docs/tutorial-basics/tutorial_02_basic_ports}.

\bibitemdeclare{inproceedings}{berthomieu2008ERTS2}
\bibitem{berthomieu2008ERTS2}
\bibinfo{author}{Bernard \surnamestart Berthomieu\surnameend},
  \bibinfo{author}{Jean-Paul \surnamestart Bodeveix\surnameend},
  \bibinfo{author}{Patrick \surnamestart Farail\surnameend},
  \bibinfo{author}{M~\surnamestart Filali\surnameend}, \bibinfo{author}{Hubert
  \surnamestart Garavel\surnameend}, \bibinfo{author}{Pierre \surnamestart
  Gaufillet\surnameend}, \bibinfo{author}{Frederic \surnamestart
  Lang\surnameend} \& \bibinfo{author}{Fran{\c c}ois \surnamestart
  Vernadat\surnameend} (\bibinfo{year}{2008}): \emph{\bibinfo{title}{{Fiacre:
  an Intermediate Language for Model Verification in the Topcased
  Environment}}}.
\newblock In: {\slshape \bibinfo{booktitle}{{4th European Congress ERTS
  Embedded Real Time Software (ERTS 2008)}}}, \bibinfo{organization}{{SEE :
  Soci{\'e}t{\'e} de l'{\'e}lectricit{\'e}, de l'{\'e}lectronique et des
  technologies de l'information et de la communication and 3AF : Association
  A{\'e}ronautique et Astronautique de France}}, \bibinfo{address}{Toulouse,
  France}, pp. \bibinfo{pages}{1--8}.
\newblock \urlprefix\url{https://inria.hal.science/inria-00262442}.

\bibitemdeclare{inproceedings}{beyer2009FMCAD}
\bibitem{beyer2009FMCAD}
\bibinfo{author}{Dirk \surnamestart Beyer\surnameend},
  \bibinfo{author}{Alessandro \surnamestart Cimatti\surnameend},
  \bibinfo{author}{Alberto \surnamestart Griggio\surnameend},
  \bibinfo{author}{M~Erkan \surnamestart Keremoglu\surnameend} \&
  \bibinfo{author}{Roberto \surnamestart Sebastiani\surnameend}
  (\bibinfo{year}{2009}): \emph{\bibinfo{title}{Software model checking via
  large-block encoding}}.
\newblock In: {\slshape \bibinfo{booktitle}{2009 Formal Methods in
  Computer-Aided Design}}, \bibinfo{organization}{IEEE}, pp.
  \bibinfo{pages}{25--32}, \doi{10.1109/FMCAD.2009.5351147}.

\bibitemdeclare{article}{biggar2020IEEE-RAL}
\bibitem{biggar2020IEEE-RAL}
\bibinfo{author}{Oliver \surnamestart Biggar\surnameend} \&
  \bibinfo{author}{Mohammad \surnamestart Zamani\surnameend}
  (\bibinfo{year}{2020}): \emph{\bibinfo{title}{A Framework for Formal
  Verification of Behavior Trees With Linear Temporal Logic}}.
\newblock {\slshape \bibinfo{journal}{IEEE Robotics and Automation Letters}}
  \bibinfo{volume}{5}(\bibinfo{number}{2}), pp. \bibinfo{pages}{2341--2348},
  \doi{10.1109/LRA.2020.2970634}.

\bibitemdeclare{misc}{btcompiler2019}
\bibitem{btcompiler2019}
\bibinfo{author}{\surnamestart CARVE-ROBMOSYS\surnameend}
  (\bibinfo{year}{2019}): \emph{\bibinfo{title}{BTCompiler}}.
\newblock \urlprefix\url{https://github.com/CARVE-ROBMOSYS/BTCompiler}.
\newblock \bibinfo{note}{Github repository}.

\bibitemdeclare{inproceedings}{nuXmv}
\bibitem{nuXmv}
\bibinfo{author}{Roberto \surnamestart Cavada\surnameend},
  \bibinfo{author}{Alessandro \surnamestart Cimatti\surnameend},
  \bibinfo{author}{Michele \surnamestart Dorigatti\surnameend},
  \bibinfo{author}{Alberto \surnamestart Griggio\surnameend},
  \bibinfo{author}{Alessandro \surnamestart Mariotti\surnameend},
  \bibinfo{author}{Andrea \surnamestart Micheli\surnameend},
  \bibinfo{author}{Sergio \surnamestart Mover\surnameend},
  \bibinfo{author}{Marco \surnamestart Roveri\surnameend} \&
  \bibinfo{author}{Stefano \surnamestart Tonetta\surnameend}
  (\bibinfo{year}{2014}): \emph{\bibinfo{title}{The nuXmv Symbolic Model
  Checker}}.
\newblock In: {\slshape \bibinfo{booktitle}{CAV}}, pp.
  \bibinfo{pages}{334--342}, \doi{10.1007/978-3-319-08867-9_22}.


\bibitemdeclare{inproceedings}{clarke2004TACAS}
\bibitem{clarke2004TACAS}
\bibinfo{author}{Edmund \surnamestart Clarke\surnameend},
  \bibinfo{author}{Daniel \surnamestart Kroening\surnameend} \&
  \bibinfo{author}{Flavio \surnamestart Lerda\surnameend}
  (\bibinfo{year}{2004}): \emph{\bibinfo{title}{A tool for checking ANSI-C
  programs}}.
\newblock In: {\slshape \bibinfo{booktitle}{International Conference on Tools
  and Algorithms for the Construction and Analysis of Systems}},
  \bibinfo{organization}{Springer}, pp. \bibinfo{pages}{168--176},
  \doi{10.1007/978-3-540-24730-2_15}.

\bibitemdeclare{inproceedings}{colledanchise2021IROS}
\bibitem{colledanchise2021IROS}
\bibinfo{author}{Michele \surnamestart Colledanchise\surnameend},
  \bibinfo{author}{Giuseppe \surnamestart Cicala\surnameend},
  \bibinfo{author}{Daniele~E. \surnamestart Domenichelli\surnameend},
  \bibinfo{author}{Lorenzo \surnamestart Natale\surnameend} \&
  \bibinfo{author}{Armando \surnamestart Tacchella\surnameend}
  (\bibinfo{year}{2021}): \emph{\bibinfo{title}{Formalizing the Execution
  Context of Behavior Trees for Runtime Verification of Deliberative
  Policies}}.
\newblock In: {\slshape \bibinfo{booktitle}{2021 IEEE/RSJ International
  Conference on Intelligent Robots and Systems (IROS)}},
  \bibinfo{publisher}{IEEE Press}, pp. \bibinfo{pages}{9841--9848},
  \doi{10.1109/IROS51168.2021.9636129}.

\bibitemdeclare{article}{cytron1991TOPLAS}
\bibitem{cytron1991TOPLAS}
\bibinfo{author}{Ron \surnamestart Cytron\surnameend}, \bibinfo{author}{Jeanne
  \surnamestart Ferrante\surnameend}, \bibinfo{author}{Barry~K \surnamestart
  Rosen\surnameend}, \bibinfo{author}{Mark~N \surnamestart Wegman\surnameend}
  \& \bibinfo{author}{F~Kenneth \surnamestart Zadeck\surnameend}
  (\bibinfo{year}{1991}): \emph{\bibinfo{title}{Efficiently computing static
  single assignment form and the control dependence graph}}.
\newblock {\slshape \bibinfo{journal}{ACM Transactions on Programming Languages
  and Systems (TOPLAS)}} \bibinfo{volume}{13}(\bibinfo{number}{4}), pp.
  \bibinfo{pages}{451--490}, \doi{10.1145/115372.115320}.

\bibitemdeclare{article}{dalzilio2023RAS}
\bibitem{dalzilio2023RAS}
\bibinfo{author}{Silvano \surnamestart Dal~Zilio\surnameend},
  \bibinfo{author}{Pierre-Emmanuel \surnamestart Hladik\surnameend},
  \bibinfo{author}{F{\'e}lix \surnamestart Ingrand\surnameend} \&
  \bibinfo{author}{Anthony \surnamestart Mallet\surnameend}
  (\bibinfo{year}{2023}): \emph{\bibinfo{title}{A formal toolchain for offline
  and run-time verification of robotic systems}}.
\newblock {\slshape \bibinfo{journal}{Robotics and Autonomous Systems}}
  \bibinfo{volume}{159}, p. \bibinfo{pages}{104301},
  \doi{10.1016/j.robot.2022.104301}.

\bibitemdeclare{article}{textX}
\bibitem{textX}
\bibinfo{author}{I.~\surnamestart Dejanović\surnameend},
  \bibinfo{author}{R.~\surnamestart Vaderna\surnameend},
  \bibinfo{author}{G.~\surnamestart Milosavljević\surnameend} \&
  \bibinfo{author}{Z.~\surnamestart Vuković\surnameend}
  (\bibinfo{year}{2017}): \emph{\bibinfo{title}{TextX: A Python tool for
  Domain-Specific Languages implementation}}.
\newblock {\slshape \bibinfo{journal}{Knowledge-Based Systems}}
  \bibinfo{volume}{115}, pp. \bibinfo{pages}{1--4},
  \doi{10.1016/j.knosys.2016.10.023}.
\newblock
  \urlprefix\url{https://www.sciencedirect.com/science/article/pii/S0950705116304178}.

\bibitemdeclare{article}{grunske2008Journal-of-Visual-Language-and-Computing}
\bibitem{grunske2008Journal-of-Visual-Language-and-Computing}
\bibinfo{author}{Lars \surnamestart Grunske\surnameend},
  \bibinfo{author}{Kirsten \surnamestart Winter\surnameend} \&
  \bibinfo{author}{Nisansala \surnamestart Yatapanage\surnameend}
  (\bibinfo{year}{2008}): \emph{\bibinfo{title}{Defining the abstract syntax of
  visual languages with advanced graph grammars—A case study based on
  behavior trees}}.
\newblock {\slshape \bibinfo{journal}{Journal of Visual Languages and
  Computing}} \bibinfo{volume}{19}(\bibinfo{number}{3}), pp.
  \bibinfo{pages}{343--379}, \doi{10.1016/j.jvlc.2007.11.003}.

\bibitemdeclare{inproceedings}{hallen2024CASE}
\bibitem{hallen2024CASE}
\bibinfo{author}{Mattias \surnamestart Hallen\surnameend},
  \bibinfo{author}{Matteo \surnamestart Iovino\surnameend},
  \bibinfo{author}{Shiva \surnamestart Sander-Tavallaey\surnameend} \&
  \bibinfo{author}{Christian \surnamestart Smith\surnameend}
  (\bibinfo{year}{2024}): \emph{\bibinfo{title}{Behavior Trees in Industrial
  Applications: A Case Study in Underground Explosive Charging}}.
\newblock In: {\slshape \bibinfo{booktitle}{2024 IEEE 20th International
  Conference on Automation Science and Engineering (CASE)}}, pp.
  \bibinfo{pages}{156--162}, \doi{10.1109/CASE59546.2024.10711822}.

\bibitemdeclare{inproceedings}{henn2022FMICS}
\bibitem{henn2022FMICS}
\bibinfo{author}{Thomas \surnamestart Henn\surnameend}, \bibinfo{author}{Marcus
  \surnamestart V{\"o}lker\surnameend}, \bibinfo{author}{Stefan \surnamestart
  Kowalewski\surnameend}, \bibinfo{author}{Minh \surnamestart
  Trinh\surnameend}, \bibinfo{author}{Oliver \surnamestart Petrovic\surnameend}
  \& \bibinfo{author}{Christian \surnamestart Brecher\surnameend}
  (\bibinfo{year}{2022}): \emph{\bibinfo{title}{Verification of Behavior Trees
  using Linear Constrained Horn Clauses}}.
\newblock In \bibinfo{editor}{Jan~Friso \surnamestart Groote\surnameend} \&
  \bibinfo{editor}{Marieke \surnamestart Huisman\surnameend}, editors:
  {\slshape \bibinfo{booktitle}{Formal Methods for Industrial Critical
  Systems}}, \bibinfo{publisher}{Springer International Publishing},
  \bibinfo{address}{Cham}, pp. \bibinfo{pages}{211--225},
  \doi{10.1007/978-3-031-15008-1_14}.

\bibitemdeclare{article}{hladik2021JSS}
\bibitem{hladik2021JSS}
\bibinfo{author}{Pierre-Emmanuel \surnamestart Hladik\surnameend},
  \bibinfo{author}{F{\'e}lix \surnamestart Ingrand\surnameend},
  \bibinfo{author}{Silvano \surnamestart Dal~Zilio\surnameend} \&
  \bibinfo{author}{Reyyan \surnamestart Tekin\surnameend}
  (\bibinfo{year}{2021}): \emph{\bibinfo{title}{Hippo: A formal-model execution
  engine to control and verify critical real-time systems}}.
\newblock {\slshape \bibinfo{journal}{Journal of Systems and Software}}
  \bibinfo{volume}{181}, p. \bibinfo{pages}{111033},
  \doi{10.1016/j.jss.2021.111033}.

\bibitemdeclare{inproceedings}{hong2023}
\bibitem{hong2023}
\bibinfo{author}{Weijiang \surnamestart Hong\surnameend},
  \bibinfo{author}{Zhenbang \surnamestart Chen\surnameend},
  \bibinfo{author}{Minglong \surnamestart Li\surnameend},
  \bibinfo{author}{Yuhan \surnamestart Li\surnameend}, \bibinfo{author}{Peishan
  \surnamestart Huang\surnameend} \& \bibinfo{author}{Ji~\surnamestart
  Wang\surnameend} (\bibinfo{year}{2024}): \emph{\bibinfo{title}{Formal
  Verification Based Synthesis for Behavior Trees}}.
\newblock In \bibinfo{editor}{Holger \surnamestart Hermanns\surnameend},
  \bibinfo{editor}{Jun \surnamestart Sun\surnameend} \& \bibinfo{editor}{Lei
  \surnamestart Bu\surnameend}, editors: {\slshape
  \bibinfo{booktitle}{Dependable Software Engineering. Theories, Tools, and
  Applications}}, \bibinfo{publisher}{Springer Nature Singapore},
  \bibinfo{address}{Singapore}, pp. \bibinfo{pages}{72--91},
  \doi{10.1007/978-981-99-8664-4_5}.

\bibitemdeclare{misc}{ingrand2025arXiv}
\bibitem{ingrand2025arXiv}
\bibinfo{author}{Felix \surnamestart Ingrand\surnameend}
  (\bibinfo{year}{2025}): \emph{\bibinfo{title}{A formal implementation of
  Behavior Trees to act in robotics}}, \doi{10.48550/arXiv.2502.11904}.
\newblock \eprint{2502.11904}.

\bibitemdeclare{article}{iovino2022Robotics-and-Autonomous-Systems}
\bibitem{iovino2022Robotics-and-Autonomous-Systems}
\bibinfo{author}{Matteo \surnamestart Iovino\surnameend},
  \bibinfo{author}{Edvards \surnamestart Scukins\surnameend},
  \bibinfo{author}{Jonathan \surnamestart Styrud\surnameend},
  \bibinfo{author}{Petter \surnamestart Ögren\surnameend} \&
  \bibinfo{author}{Christian \surnamestart Smith\surnameend}
  (\bibinfo{year}{2022}): \emph{\bibinfo{title}{A survey of Behavior Trees in
  robotics and AI}}.
\newblock {\slshape \bibinfo{journal}{Robotics and Autonomous Systems}}
  \bibinfo{volume}{154}, p. \bibinfo{pages}{104096},
  \doi{10.1016/j.robot.2022.104096}.

\bibitemdeclare{article}{ROS2}
\bibitem{ROS2}
\bibinfo{author}{Steven \surnamestart Macenski\surnameend},
  \bibinfo{author}{Tully \surnamestart Foote\surnameend},
  \bibinfo{author}{Brian \surnamestart Gerkey\surnameend},
  \bibinfo{author}{Chris \surnamestart Lalancette\surnameend} \&
  \bibinfo{author}{William \surnamestart Woodall\surnameend}
  (\bibinfo{year}{2022}): \emph{\bibinfo{title}{Robot Operating System 2:
  Design, architecture, and uses in the wild}}.
\newblock {\slshape \bibinfo{journal}{Science Robotics}}
  \bibinfo{volume}{7}(\bibinfo{number}{66}), p. \bibinfo{pages}{eabm6074},
  \doi{10.1126/scirobotics.abm6074}.
\newblock
  \urlprefix\url{https://www.science.org/doi/abs/10.1126/scirobotics.abm6074}.

\bibitemdeclare{misc}{MoVe4BT}
\bibitem{MoVe4BT}
\bibinfo{author}{Huang \surnamestart Peishan\surnameend}, \bibinfo{author}{Hong
  \surnamestart Weijiang\surnameend}, \bibinfo{author}{Chen \surnamestart
  Zhenbang\surnameend} \& \bibinfo{author}{Wang \surnamestart Ji\surnameend}:
  \emph{\bibinfo{title}{MoVe4BT: Modeling \& Verification For BT}}.
\newblock \urlprefix\url{https://move4bt.github.io/}.
\newblock \bibinfo{note}{Accessed: 2023-12-14}.

\bibitemdeclare{inproceedings}{potteiger2024ICAA}
\bibitem{potteiger2024ICAA}
\bibinfo{author}{Nicholas \surnamestart Potteiger\surnameend},
  \bibinfo{author}{Ankita \surnamestart Samaddar\surnameend},
  \bibinfo{author}{Hunter \surnamestart Bergstrom\surnameend} \&
  \bibinfo{author}{Xenofon \surnamestart Koutsoukos\surnameend}
  (\bibinfo{year}{2024}): \emph{\bibinfo{title}{Designing Robust Cyber-Defense
  Agents with Evolving Behavior Trees}}.
\newblock In: {\slshape \bibinfo{booktitle}{2024 International Conference on
  Assured Autonomy (ICAA)}}, pp. \bibinfo{pages}{1--10},
  \doi{10.1109/ICAA64256.2024.00011}.

\bibitemdeclare{inproceedings}{rocamora2024ICUAS}
\bibitem{rocamora2024ICUAS}
\bibinfo{author}{Bernardo~Martinez \surnamestart Rocamora\surnameend},
  \bibinfo{author}{Paulo V.~G. \surnamestart Simplicio\surnameend} \&
  \bibinfo{author}{Guilherme A.~S. \surnamestart Pereira\surnameend}
  (\bibinfo{year}{2024}): \emph{\bibinfo{title}{A Behavior Tree Approach for
  Battery-Aware Inspection of Large Structures Using Drones}}.
\newblock In: {\slshape \bibinfo{booktitle}{2024 International Conference on
  Unmanned Aircraft Systems (ICUAS)}}, pp. \bibinfo{pages}{234--240},
  \doi{10.1109/ICUAS60882.2024.10557083}.

\bibitemdeclare{inproceedings}{samaddar2025ICAIC}
\bibitem{samaddar2025ICAIC}
\bibinfo{author}{Ankita \surnamestart Samaddar\surnameend},
  \bibinfo{author}{Nicholas \surnamestart Potteiger\surnameend} \&
  \bibinfo{author}{Xenofon \surnamestart Koutsoukos\surnameend}
  (\bibinfo{year}{2025}): \emph{\bibinfo{title}{Out-of-Distribution Detection
  for Neurosymbolic Autonomous Cyber Agents}}.
\newblock In: {\slshape \bibinfo{booktitle}{2025 IEEE 4th International
  Conference on AI in Cybersecurity (ICAIC)}}, pp. \bibinfo{pages}{1--9},
  \doi{10.1109/ICAIC63015.2025.10849024}.

\bibitemdeclare{article}{scheide2025AutonomousRobots}
\bibitem{scheide2025AutonomousRobots}
\bibinfo{author}{Emily \surnamestart Scheide\surnameend},
  \bibinfo{author}{Graeme \surnamestart Best\surnameend} \&
  \bibinfo{author}{Geoffrey~A. \surnamestart Hollinger\surnameend}
  (\bibinfo{year}{2025}): \emph{\bibinfo{title}{Synthesizing compact behavior
  trees for probabilistic robotics domains}}.
\newblock {\slshape \bibinfo{journal}{Autonomous Robots}}
  \bibinfo{volume}{49}(\bibinfo{number}{1}), p.~\bibinfo{pages}{3},
  \doi{10.1007/s10514-024-10187-z}.

\bibitemdeclare{inproceedings}{serbinowska2022SEFM}
\bibitem{serbinowska2022SEFM}
\bibinfo{author}{Serena~S. \surnamestart Serbinowska\surnameend} \&
  \bibinfo{author}{Taylor~T. \surnamestart Johnson\surnameend}
  (\bibinfo{year}{2022}): \emph{\bibinfo{title}{BehaVerify: Verifying Temporal
  Logic Specifications For Behavior Trees}}.
\newblock In: {\slshape \bibinfo{booktitle}{Software Engineering and Formal
  Methods: 20th International Conference, SEFM 2022, Berlin, Germany, September
  26-30, 2022, Proceedings}}, \bibinfo{publisher}{Springer-Verlag},
  \bibinfo{address}{Berlin, Heidelberg}, pp. \bibinfo{pages}{307--323},
  \doi{10.1007/978-3-031-17108-6_19}.

\bibitemdeclare{inproceedings}{serbinowska2025NEUS-REPLACE}
\bibitem{serbinowska2025NEUS-REPLACE}
\bibinfo{author}{Serena~S. \surnamestart Serbinowska\surnameend},
  \bibinfo{author}{Diego \surnamestart Manzanas~Lopez\surnameend},
  \bibinfo{author}{Dung~Thuy \surnamestart Nguyen\surnameend} \&
  \bibinfo{author}{Taylor~T. \surnamestart Johnson\surnameend}
  (\bibinfo{year}{2025}): \emph{\bibinfo{title}{Neuro-Symbolic Behavior Trees
  (NSBTs) and Their Verification}}.
\newblock In: {\slshape \bibinfo{booktitle}{Proceedings of the International
  Conference on Neuro-symbolic Systems}}, {\slshape
  \bibinfo{series}{Proceedings of Machine Learning Research}}
  \bibinfo{volume}{288}, \bibinfo{publisher}{PMLR}, pp.
  \bibinfo{pages}{409--423}.
\newblock
  \urlprefix\url{https://proceedings.mlr.press/v288/serbinowska25a.html}.

\bibitemdeclare{inproceedings}{serbinowska2024FMAS-BTM}
\bibitem{serbinowska2024FMAS-BTM}
\bibinfo{author}{Serena~S. \surnamestart Serbinowska\surnameend},
  \bibinfo{author}{Nicholas \surnamestart Potteiger\surnameend},
  \bibinfo{author}{Anne~M. \surnamestart Tumlin\surnameend} \&
  \bibinfo{author}{Taylor~T. \surnamestart Johnson\surnameend}
  (\bibinfo{year}{2024}): \emph{\bibinfo{title}{Verification of Behavior Trees
  with Contingency Monitors}}.
\newblock In \bibinfo{editor}{Matt \surnamestart Luckcuck\surnameend} \&
  \bibinfo{editor}{Mengwei \surnamestart Xu\surnameend}, editors: {\slshape
  \bibinfo{booktitle}{{\rm Proceedings Sixth International Workshop on} Formal
  Methods for Autonomous Systems, {\rm Manchester, UK, 11th and 12th of
  November 2024}}}, {\slshape \bibinfo{series}{Electronic Proceedings in
  Theoretical Computer Science}} \bibinfo{volume}{411},
  \bibinfo{publisher}{Open Publishing Association}, pp.
  \bibinfo{pages}{56--72}, \doi{10.4204/EPTCS.411.4}.

\bibitemdeclare{inproceedings}{serbinowska2024FMAS-SBT}
\bibitem{serbinowska2024FMAS-SBT}
\bibinfo{author}{Serena~S. \surnamestart Serbinowska\surnameend},
  \bibinfo{author}{Preston \surnamestart Robinette\surnameend},
  \bibinfo{author}{Gabor \surnamestart Karsai\surnameend} \&
  \bibinfo{author}{Taylor~T. \surnamestart Johnson\surnameend}
  (\bibinfo{year}{2024}): \emph{\bibinfo{title}{Formalizing Stateful Behavior
  Trees}}.
\newblock In \bibinfo{editor}{Matt \surnamestart Luckcuck\surnameend} \&
  \bibinfo{editor}{Mengwei \surnamestart Xu\surnameend}, editors: {\slshape
  \bibinfo{booktitle}{{\rm Proceedings Sixth International Workshop on} Formal
  Methods for Autonomous Systems, {\rm Manchester, UK, 11th and 12th of
  November 2024}}}, {\slshape \bibinfo{series}{Electronic Proceedings in
  Theoretical Computer Science}} \bibinfo{volume}{411},
  \bibinfo{publisher}{Open Publishing Association}, pp.
  \bibinfo{pages}{201--218}, \doi{10.4204/EPTCS.411.14}.

\bibitemdeclare{misc}{behaverify2026zenodo}
\bibitem{behaverify2026zenodo}
\bibinfo{author}{Serena~Serafina \surnamestart Serbinowska\surnameend},
  \bibinfo{author}{Ben \surnamestart Wooding\surnameend},
  \bibinfo{author}{Diego \surnamestart Manzanas~Lopez\surnameend},
  \bibinfo{author}{Luke \surnamestart Nam\surnameend}, \bibinfo{author}{Anne
  \surnamestart Tumlin\surnameend}, \bibinfo{author}{Nicholas~K. \surnamestart
  Potteiger\surnameend}, \bibinfo{author}{Hongchao \surnamestart
  Zhang\surnameend} \& \bibinfo{author}{Taylor \surnamestart
  Johnson\surnameend} (\bibinfo{year}{2026}):
  \emph{\bibinfo{title}{{BehaVerify}: A Formal Verification Tool for Behavior
  Trees (software, version v1.0)}}.
\newblock \bibinfo{howpublished}{Zenodo}, \doi{10.5281/zenodo.20280547}.

\bibitemdeclare{inproceedings}{shin2024ICTC}
\bibitem{shin2024ICTC}
\bibinfo{author}{Mingyu \surnamestart Shin\surnameend} \& \bibinfo{author}{Soyi
  \surnamestart Jung\surnameend} (\bibinfo{year}{2024}):
  \emph{\bibinfo{title}{A Survey of Behavior Tree-Based Task Planning
  Algorithms for Autonomous Robotic Systems}}.
\newblock In: {\slshape \bibinfo{booktitle}{2024 15th International Conference
  on Information and Communication Technology Convergence (ICTC)}}, pp.
  \bibinfo{pages}{2039--2041}, \doi{10.1109/ICTC62082.2024.10827191}.

\bibitemdeclare{misc}{PyTrees}
\bibitem{PyTrees}
\bibinfo{author}{Daniel \surnamestart Stonier\surnameend}:
  \emph{\bibinfo{title}{PyTrees Module API}}.
\newblock
  \urlprefix\url{https://py-trees.readthedocs.io/en/devel/modules.html}.
\newblock \bibinfo{note}{Accessed: 2023-12-14}.

\bibitemdeclare{manual}{Coq}
\bibitem{Coq}
\bibinfo{author}{The Coq~Development \surnamestart Team\surnameend}
  (\bibinfo{year}{2024}): \emph{\bibinfo{title}{The Coq Proof Assistant,
  Reference Manual}}.
\newblock \urlprefix\url{https://coq.inria.fr/doc/V8.20.0/refman/}.

\bibitemdeclare{inproceedings}{wang2024FACS}
\bibitem{wang2024FACS}
\bibinfo{author}{Qiang \surnamestart Wang\surnameend}, \bibinfo{author}{Huadong
  \surnamestart Dai\surnameend}, \bibinfo{author}{Yongxin \surnamestart
  Zhao\surnameend}, \bibinfo{author}{Min \surnamestart Zhang\surnameend} \&
  \bibinfo{author}{Simon \surnamestart Bliudze\surnameend}
  (\bibinfo{year}{2024}): \emph{\bibinfo{title}{Enabling Behaviour Tree
  Verification via a Translation to BIP}}.
\newblock In \bibinfo{editor}{Diego \surnamestart Marmsoler\surnameend} \&
  \bibinfo{editor}{Meng \surnamestart Sun\surnameend}, editors: {\slshape
  \bibinfo{booktitle}{Formal Aspects of Component Software}},
  \bibinfo{publisher}{Springer Nature Switzerland}, \bibinfo{address}{Cham},
  pp. \bibinfo{pages}{3--20}, \doi{10.1007/978-3-031-71261-6_1}.

\bibitemdeclare{inproceedings}{wooding2026selfcorrecting}
\bibitem{wooding2026selfcorrecting}
\bibinfo{author}{Ben \surnamestart Wooding\surnameend},
  \bibinfo{author}{Kiersten \surnamestart Brennan\surnameend},
  \bibinfo{author}{Anne~M \surnamestart Tumlin\surnameend},
  \bibinfo{author}{Hongchao \surnamestart Zhang\surnameend} \&
  \bibinfo{author}{Taylor~T \surnamestart Johnson\surnameend}
  (\bibinfo{year}{2026}): \emph{\bibinfo{title}{A Self-correcting
  Neuro-symbolic AI Reasoning Framework}}.
\newblock In: {\slshape \bibinfo{booktitle}{International Symposium on AI
  Verification}}, \bibinfo{organization}{Springer}, pp.
  \bibinfo{pages}{191--211}, \doi{10.1007/978-3-032-32357-6_9}.

\bibitemdeclare{inproceedings}{wu2024ICRA}
\bibitem{wu2024ICRA}
\bibinfo{author}{Yansong \surnamestart Wu\surnameend}, \bibinfo{author}{Fan
  \surnamestart Wu\surnameend}, \bibinfo{author}{Lingyun \surnamestart
  Chen\surnameend}, \bibinfo{author}{Kejia \surnamestart Chen\surnameend},
  \bibinfo{author}{Samuel \surnamestart Schneider\surnameend},
  \bibinfo{author}{Lars \surnamestart Johannsmeier\surnameend},
  \bibinfo{author}{Zhenshan \surnamestart Bing\surnameend},
  \bibinfo{author}{Fares~J. \surnamestart Abu-Dakka\surnameend},
  \bibinfo{author}{Alois \surnamestart Knoll\surnameend} \&
  \bibinfo{author}{Sami \surnamestart Haddadin\surnameend}
  (\bibinfo{year}{2024}): \emph{\bibinfo{title}{1 kHz Behavior Tree for
  Self-adaptable Tactile Insertion}}.
\newblock In: {\slshape \bibinfo{booktitle}{2024 IEEE International Conference
  on Robotics and Automation (ICRA)}}, pp. \bibinfo{pages}{16002--16008},
  \doi{10.1109/ICRA57147.2024.10610835}.

\bibitemdeclare{inproceedings}{zubrycki2024RoMoCo}
\bibitem{zubrycki2024RoMoCo}
\bibinfo{author}{Igor \surnamestart Zubrycki\surnameend}
  (\bibinfo{year}{2024}): \emph{\bibinfo{title}{Behavior Tree Based Controller
  for a Soft Rehabilitation Device Stretchbox}}.
\newblock In: {\slshape \bibinfo{booktitle}{2024 13th International Workshop on
  Robot Motion and Control (RoMoCo)}}, pp. \bibinfo{pages}{168--173},
  \doi{10.1109/RoMoCo60539.2024.10604409}.

\end{thebibliography}


\end{document}